\documentclass[12pt]{article}
\pdfoutput=1
\usepackage[pdftex]{graphicx}
\usepackage{amssymb}
\usepackage{amsmath}
\usepackage{bm}
\usepackage{enumerate}

\newcommand{\sub}[1]{_{\mbox{\tiny\emph{#1}}}}

\begin{document}

\begin{titlepage}

	\vskip 2cm
	\begin{center}
		\Large{{\bf Emergent geometry from stochastic dynamics,\\or Hawking evaporation in M(atrix) theory}}
	\end{center}

	\vskip 2cm
	\begin{center}
		{Haoxing Du\footnote{\tt{hdu@g.hmc.edu}} and Vatche Sahakian\footnote{\tt{sahakian@hmc.edu}}}\\
	\end{center}
	\vskip 12pt
	\centerline{\sl Harvey Mudd College}
	\centerline{\sl Physics Department, 241 Platt Blvd.}
	\centerline{\sl Claremont CA 91711 USA}

	\vskip 1cm
	\begin{abstract}
		We develop an microscopic model of the M-theory Schwarzschild black hole using the Banks-Fischler-Shenker-Susskind Matrix formulation of quantum gravity. The underlying dynamics is known to be chaotic, which allows us to use methods from Random Matrix Theory and non-equilibrium statistical mechanics to propose a coarse-grained bottom-up picture of the event horizon -- and the associated Hawking evaporation phenomenon. The analysis is possible due to a hierarchy between the various timescales at work. Event horizon physics is found to be non-local 
at the Planck scale, and we demonstrate how non-unitary physics and information loss arise from the process of averaging over the chaotic unitary dynamics. Most interestingly, we correlate the onset of non-unitarity with the emergence of spacetime geometry outside the horizon. We also 
write a mean field action for the evolution of qubits -- represented by polarization states of supergravity modes. This evolution is shown to have similarities to a recent toy model of black hole evaporation proposed by Osuga and Page -- a model aimed at developing a plausible no-firewall scenario.
	\end{abstract}
\end{titlepage}

\newpage \setcounter{page}{1}

\section{Introduction and highlights}
\label{sub:intro}

The study of black holes in M(atrix) theory holds a treasure trove of insight into quantum gravity and the nature of spacetime. As a non-perturbative formulation of M-theory, Matrix theory~\cite{Banks:1996vh,Bigatti:1997jy} can in principle access and potentially resolve many of the puzzles we associate with black holes. Early attempts at staging Matrix black holes have consisted of promising sketches~\cite{Horowitz:1997fr}-\cite{Berkowitz:2016znt} and numerical simulations~\cite{Hanada:2010rg}-\cite{Berkowitz:2016muc}. We have learned that understanding black holes is related to studying strongly coupled Yang-Mills at finite temperature~\cite{Itzhaki:1998dd}-\cite{Martinec:1998ja}, and that there might be intricate non-local dynamics near the event horizon~\cite{Giddings:2011ks,Giddings:2012bm}. More recently, we have learned that Matrix theory is characteristically chaotic~\cite{Berkowitz:2016znt,Maldacena:2015waa,Gur-Ari:2015rcq}, and interactions can scramble initial value data at the fastest possible rate that is allowed by the postulates of quantum mechanics~\cite{Sekino:2008he}-\cite{Gharibyan:2018jrp} -- as also expected from black hole physics. 

In this work we ask if one can write a mean field coarse-grained description of the strongly coupled microscopic dynamics of Matrix theory in a manner that captures the essential features of black holes and informs us about the geometry near the event horizon. To illustrate through an analogy, if M(atrix) theory is to black hole quantum mechanics as BCS theory is to superconductivity, we are looking for the analogue of a Landau-Ginzburg description of the quantum physics of black holes -- with the underpinning element of stochastic chaotic evolution.

We know that Matrix theory is chaotic, and we know that one can often use the language of random variables, or in this case Random Matrix theory (RMT)~\cite{Gharibyan:2018jrp}-\cite{Cotler:2016fpe}\cite{Berkowitz:2016znt}, to capture chaotic dynamics. We also know that RMT is closely related to the strong damping regime of Fokker-Planck stochastic evolution~\cite{fokker,lemons,mahnke,Dyson1} whereby a statistical description of ergodic motion is effectively described with macroscopic variables. The suggestion is then to formulate a description of Matrix black holes where the entries of the Matrices are described through particles moving in a mean field potential -- one that is obtained by coarse-graining over microscopic degrees of freedom that are engaged in ergodic motion.

\begin{figure}
	\begin{center}
		\includegraphics[width=2.5in]{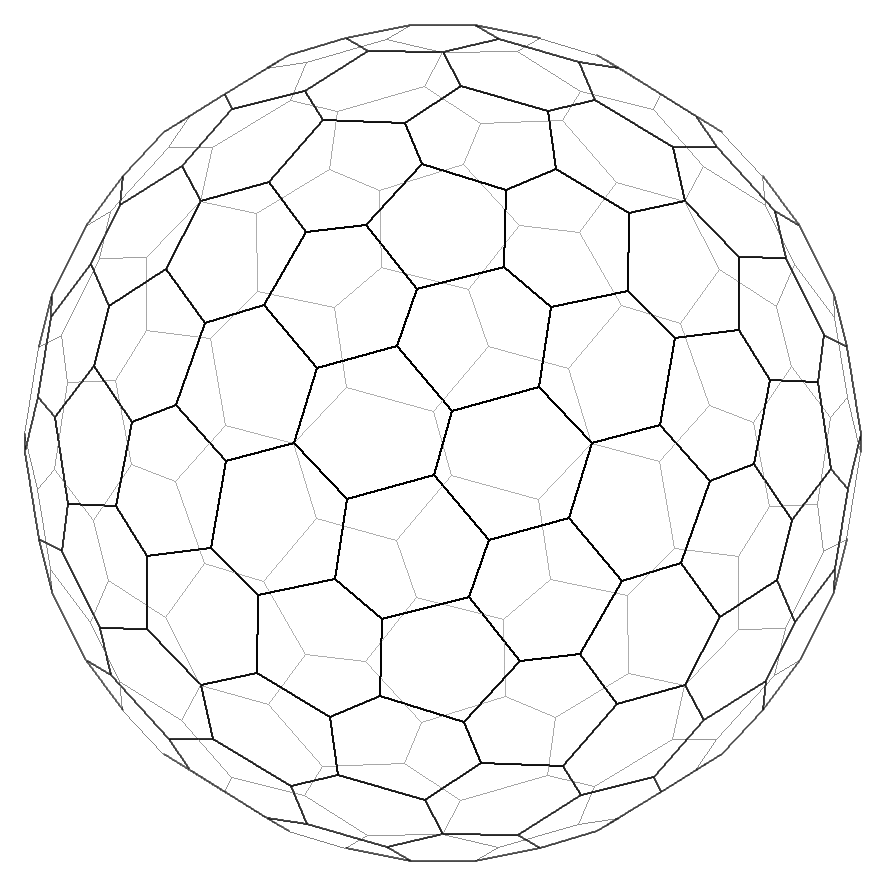}
	\end{center}
	\caption{A cartoon of the effective model of the light-cone Schwarzschild black hole. The
cells represent Planck size  marginally bound D0 branes, about $d$ per cell in $d$ space dimensions. The cells are glued together with a condensate of off-diagonal matrix modes that act as scaffolding and do not carry information or entropy.}\label{fig:bh}
\end{figure}
In this work, we show that such an effective description of black holes is indeed possible using Matrix theory. In the process of developing this effective model, we settle on a microscopic picture of Matrix black holes that is both intuitive and complex. Entries on the diagonal of the matrices incorporate the thermodynamics and encode information. These can be thought of as particles that mostly hang around near the surface of the would-be horizon. They are subject to a mean field potential whose shape we determine. An additional `goo' of off-diagonal matrix
entries glue these particles into clusters, effectively acting like bound states. These clusters contain around $d$ particles each, for a black hole in $d$ space dimensions. Figure~\ref{fig:bh} depicts a cartoon of the model. In the figure, the clusters are depicted as cells. The configuration is far from static, and in fact we expect that the cells continuously exchange particles and rearrange themselves. The rest of the matrix degrees of freedom, which constitute the overwhelming majority of the total, condense in a quantum ground state. It is possible that they should be thought of as a membrane stretched at the horizon, without any associated thermodynamics or entropy. Thermal energy is distributed in the dynamics of the cells as they slide near the horizon and interact with each other.

We develop this model in detail, matching with all expectations from the dual M-theory supergravity description of a Schwarzschild black hole in the light-cone frame. In particular, Hawking evaporation~\cite{Hawking:1974sw}-\cite{Majhi:2011yi} is reproduced and information loss is demonstrated to arise from the process of coarse-graining over otherwise unitary dynamics. It
becomes clear that dynamics near the horizon has a non-local component when explored at short enough timescales, while being local at the longer timescales associated with Hawking radiation\footnote{To clarify, this non-locality arises at the Planck scale. At energy scales below the Planck scale, we see no evidence for non-locality. This is the same non-local phenomenon typically associated with D0 brane scattering.}. Most interestingly, we demonstrate that non-unitary evolution and information loss arise at the timescales for which the Matrix dynamics is strongly coupled and spacetime geometry is expected to be emergent in the dual supergravity language. This suggests that Hawking information loss is inherently tied to the premise that geometry near the horizon of a large black hole is smooth and well-defined. The microscopic degrees of freedom underlying black hole dynamics are Planck sized bits that are 
interacting chaotically over Planckian timescales. Any description of the physics over timescales larger than the Planck time involves coarse graining over stochastic dynamics in a manner that leads to an effective quantum picture that is non-unitary. The notion of spacetime geometry arises at around those Planckian timescales, implying the breakdown of the geometrical picture of black hole evaporation as we approach the horizon. Put differently, the Hawking computation is robust when applied in smooth spacetime backgrounds over large enough timescales, yet the evaporation should still be regarded as unitary because the notion of geometry and spacetime is lost at the event horizon {\em at short timescales}. 

The outline of the text is as follows. In the first section, we present a brief overview of Matrix theory, followed by a review of Fokker-Planck dynamics and the light-cone Schwarzschild black hole in supergravity. We then systematically develop the effective model for the Matrix black hole, matching and checking against expectations on the dual low energy M-theory side. In the second section, we focus on the time evolution of information within the Matrix black hole. We track information encoded in the polarization states of the low energy M-theory supergravity multiplet, and we write an effective qubit time evolution operator that is based on the stochastic model developed earlier. We show how the evolution becomes non-unitary at longer timescales because of the coarse-graining over chaotic dynamics, and correlate this with the emergence of spacetime geometry in the dual M-theory language. For short timescales, we write a unitary time evolution operator that describe the weakly coupled qubit dynamics near the event horizon. Finally, in the discussion section, we reflect on the implications and future directions.

\section{The effective model}\label{sec:first}

\subsection{M(atrix) theory overview}

The M(atrix) theory action is the dimensional reduction of $10$ dimensional Super Yang-Mills (SYM) to $0+1$ dimensions and is given by
\begin{equation}\label{eq:sym}
	S = \int dt\ \mbox{Tr}\left[
	\frac{1}{2\,R} \dot{X}_i^2+\frac{R}{2\,\lambda^3} [X_i,X_j]^2 + \frac{1}{2} \Psi \dot{\Psi} + \frac{R}{2\,\lambda^{3/2}} \Psi \Gamma^i [X_i, \Psi]
	\right]\ .
\end{equation}
The gauge group is $U(N)$, with the $X_i$s ($i=1,\ldots, 9$) and the $\Psi$ in the adjoint representation of the group. In our conventions, we have
\begin{equation}
	R = g\sub{s} \ell\sub{s}\ \ \ ,\ \ \ \lambda = 2\pi\,\ell\sub{s}^2\ ,
\end{equation} 
where $g\sub{s}$ is the string coupling and $\ell\sub{s}$ is the string length\footnote{
Matrix theory is sometimes written in Planck scale conventions, related to the one we use by $X\rightarrow Y/\sqrt{R}$ and $t\rightarrow \tau/R$.
Using units such that $2\pi\,\ell\sub{P}^3=1$ where $\ell\sub{P}$ is the eleven dimensional Planck length, the action takes the form
\begin{equation}
	S = \int d\tau\ \mbox{Tr}\left[
	\frac{1}{2\,R} \dot{Y}_i^2+\frac{R}{2} [Y_i,Y_j]^2 + \frac{1}{2} \Psi \dot{\Psi} + \frac{R}{2} \Psi \Gamma^i [Y_i, \Psi]
	\right]\ ,
\end{equation}
where $\dot{Y}=dY/d\tau$. In this alternate convention, the length dimensions of the various quantities become
$X\simeq \ell^{3/2}$, $\psi\sim \ell^0$, $t\sim \ell^2$, $R\sim \ell$. Note that 
if $Y\sim \ell\sub{P}^{3/2}$, then $X=\ell\sub{s}$, given that $\ell\sub{P} = g\sub{s}^{1/3}\ell\sub{s}$.}. The Yang-Mills coupling is
\begin{equation}
	g\sub{YM}^2 = \frac{g\sub{s}}{\ell\sub{s}^3}\ .
\end{equation}
The length dimensions of the various quantities are: $X\sim \ell^1$, $t\sim \ell^1$, and $\psi \sim \ell^0$.

The theory is purported to be a non-perturbative formulation of M-theory in the light-cone frame in the following scaling limit\footnote{This scaling limit corresponds to the decoupling regime for holographic duality~\cite{Maldacena:1997re,Itzhaki:1998dd,Witten:1998qj,Gubser:1998bc} -- as applied to D0 branes. The Matrix theory conjecture is thus in the same class of gravity-SYM correspondences that give rise to the AdS/CFT map.}
\begin{equation}
	g\sub{s},\ell\sub{s} \rightarrow 0\ \ \ \mbox{with}\ \ \ g\sub{YM}^2 = \frac{g\sub{s}}{\ell\sub{s}^3} = \mbox{fixed}\ \ \mbox{and}\ \  \frac{X}{\ell\sub{s}} = \mbox{fixed}\ .
\end{equation}
This corresponds to focusing on energies that scale as $E\sim g\sub{s}/\ell\sub{s}$. It is sometimes convenient to introduce alternate M-theory variables $\epsilon$, $\tau$, and $\xi$ that remain fixed in the scaling regime of interest
\begin{equation}\label{eq:rescaled}
	E = g\sub{s}^{2/3} \epsilon\ \ \ ,\ \ \ t = g\sub{s}^{-2/3} \tau\ \ ,\ \ X = \ell\sub{s} \xi\ .
\end{equation}
For example, the corresponding light-cone M-theory energy scale is $\epsilon \sim R/\ell\sub{P}^2 = \mbox{fixed}$. In the map onto light-cone M-theory, $N/R$ is interpreted as total light-cone momentum. Light-cone energy scales inversely with light-cone momentum, hence as $(R/N) \times \mbox{mass}^2$. Depending on the coupling regime, the number of active degrees of freedom of a configuration scales as $N^k$, where $k=2$ in the weakly coupled regime, and $k = 1$ at strong coupling. 

Compactifying light-cone M-theory to $d$ space dimensions, we can describe it through Matrix theory with $d$ of the $9$ $X_i$ matrices removed from the dynamics, assuming that the compact directions are small enough that associated modes are too heavy to excite. Alternatively, one can use $d+1$ dimensional SYM for a full description of the compactified theory, obtained from the current setup via a T-duality map. 

The relation between light-cone M-theory and Matrix theory is known to hold for $N\rightarrow \infty$, but the correspondence is valid for finite $N$ as well -- between Discrete Light-Cone Quantized (DLCQ) M-theory and finite $N$ matrix theory, where $N$ is mapped onto units of M-theory discrete light-cone momentum~\cite{Seiberg:1997ad}. In this work, we will work at finite but large $N$ in trying to describe an M-theory black hole that is large enough to have small curvature scales at its horizon.

\subsection{From chaos to a stochastic evolution}

Recently, Matrix theory has been demonstrated to be highly chaotic~\cite{Maldacena:2015waa,Gur-Ari:2015rcq,Berkowitz:2016znt}, with dynamics that can scramble initial value data in a time that scales logarithmically with the
entropy~\cite{Barbon:2011pn,Lashkari:2011yi,Brady:2013opa,Pramodh:2014jha,Gharibyan:2018jrp} -- as opposed to the more common 
power law behavior. This allows one to capture Matrix theory physics, in the appropriate setting, by treating the matrix entries as random variables. Describing a non-extremal black hole is certainly a good candidate setup for exploring chaos in Matrix theory~\cite{Cotler:2016fpe,Magan:2016ojb,Gharibyan:2018jrp}. And techniques from the well-established field of Random Matrix Theory (RMT)~\cite{Dyson1,rmt,intrormt,randommatrix,Eynard:2015aea} can then be used to tackle the problem. RMT is most powerful when one is dealing with a theory with a single matrix; it then allows a robust statistical treatment of the eigenvalues of this matrix. 

In our setup, we will be interested in studying a configuration of matrices in Matrix theory that represents a $d$ dimensional Schwarzschild black hole in the dual light-cone M-theory. We will assume from the outset that we work with spherically symmetric configurations, where the different $X_i$ matrices are chaotic and {\em uncorrelated} in different space directions. Hence, each matrix entry in the $d$ matrices $X_i$, with $i=1,\ldots, d$, is random and not correlated with any other matrix entry. This configuration is to be mapped onto a black hole in the dual M-theory -- with a fixed temperature and associated Hawking evaporation phenomenon. The fermionic matrix entries of $\Psi$ in~(\ref{eq:sym}) will be treated as a component of the thermal soup -- in equilibrium with the bosonic matrix entries. At finite temperature, we will hence mostly focus on the bosonic sector with a mirror image at play in the fermionic sector being implied. However, we do need to incorporate the one-loop quantum contribution of the 
fermionic degrees of freedom to the mean field potential for the bosonic stochastic variables. Furthermore, later on, we will use the fermionic variables as probes to track information evolution in this thermal soup. 

We start by noting that RMT is closely related to stochastic physics. In particular, since the work by Dyson~\cite{Dyson1}, it has been demonstrated that RMT dynamics can be properly captured by the strong damping regime of Fokker-Planck evolution. We present here a quick overview of the subject.

In RMT, each matrix entry can be thought of as a stochastic particle evolving in an {\em mean field potential}. For a particle with position $\bm{r}$ and velocity $\bm{v}$ in $d$ space dimensions, we can study it through the probability function
\begin{equation}
	p(\bm{r},\bm{v}, t)\ \mbox{d}^d \bm{r} \mbox{d}^d \bm{v}\ ,
\end{equation}
which represents the probability of finding the particle at time $t$ within $\bm{r}$ and $\bm{r}+\mbox{d}\bm{r}$ and  $\bm{v}$ and $\bm{v}+\mbox{d}\bm{v}$. In our setup, we will consider matrix configurations that are spherically symmetric in $d$ dimensions. We will then focus on probability profiles where
\begin{equation}
	p(\bm{r},\bm{v}, t) \rightarrow p(r, v, t) \prod_i \delta(v_{\theta_i}) \ .
\end{equation}
Here, the $v_{\theta_i}$ are $d-1$ components of $\bm{v}$ in the angular directions, and 
$v=v_r$. Correspondingly, the mean field potential is spherically symmetric\footnote{The model we develop involves time averaging over stochastic, chaotic dynamics. The cluster tiling of Figure~\ref{fig:bh} is not rigid and very dynamical over timescales shorter than the Hawking timescale. It is then reasonable to expect that, at timescales larger than the characteristic timescale associated with cluster dynamics, an approximate spherical symmetry sets in. Of course, going beyond this coarse model one needs to consider the possible breaking of the spherical symmetry~\cite{Rinaldi:2017mjl,Brower:2018szu}.}
\begin{equation}
	V(\bm{r})\rightarrow V(r)
\end{equation}
and the Fokker-Planck equation takes the form
\begin{equation}
	\frac{\partial p(r,v,t)}{\partial t} = \left(
	-v \frac{\partial}{\partial r} + \frac{1}{m} \frac{\partial V}{\partial r} \frac{\partial}{\partial v} + \gamma\, d + \gamma\, v\, \frac{\partial}{\partial v} + \frac{\gamma}{m} T \frac{1}{v^{d-1}} \frac{\partial}{\partial v} \left(v^{d-1} \frac{\partial}{\partial v}\right)
	\right) p(r,v,t)\ ,
\end{equation}
where $T$ is the temperature of the environment, $\gamma$ is a damping parameter, and $m$ is the mass of the particle. This then allows us to study the evolution of the matrix entry in a statistical framework.
The spherically symmetric Fokker-Planck equation is solved by the equilibrium time-independent profile
\begin{equation}
	p_{eq} = C\, \exp\left[-\frac{1}{T} \left(\frac{1}{2}m\,v^2+V(r)\right)\right]\prod_i \delta(v_{\theta_i})\ .
\end{equation}
$C$ here is a normalization constant. Note that this non-relativistic treatment is consistent with Matrix theory since light-cone M-theory has Galilean symmetry with dispersion relation $E\sub{LC} = \bm{p}^2/2 p\sub{LC}$, where the light-cone momentum $p\sub{LC} \sim 1/R$ plays the role of Galilean mass. 

As mentioned above, the relation between RMT and stochastic physics arises in the regime of strong damping
\begin{equation}\label{eq:strongdamping}
	\gamma \gtrsim \sqrt{-\frac{V''(0)}{m}}\ .
\end{equation}
Focusing on this regime, we also write the probability profile as
\begin{equation}
	p\rightarrow \int \mbox{d}^d \bm{v}\, p
\end{equation}
integrating over all velocities. 
The resulting evolution equation is known as the Smoluchowski equation
\begin{equation}\label{eq:smol}
	\frac{\partial p(r,t)}{\partial t} = \frac{1}{m\,\gamma}\left(
		\frac{1}{r^{d-1}} \frac{\partial}{\partial r} r^{d-1} V'(r)+\frac{T}{r^{d-1}} \frac{\partial}{\partial r} r^{d-1}\frac{\partial}{\partial r}
	\right) p(r,t)\ .
\end{equation}
The radial probability current that follows from~(\ref{eq:smol}) takes the form
\begin{equation}
	j_r = - \frac{1}{m\,\gamma} \left(T\frac{\partial}{\partial r}+V'(r)\right) p(r,t)\ ,
\end{equation}
which we will use later in understanding evaporation through stochastic diffusion. 

Our goal is to develop an effective model for strongly coupled chaotic Matrix theory, using the Smoluchowski equation with $r$ representing matrix entries in the bosonic matrix $\sqrt{\sum_i X_i^2}\sim X_i$ of~(\ref{eq:sym}) -- since different directions in space are statistically uncorrelated. We then need to identify the relevant mean field potential $V(r)$, mass $m$, temperature $T$, and damping parameter $\gamma$.  

It is worthwhile noting that an alternate and equivalent approach is to track the evolution of moments of random matrix entries. If $\chi$ represent any matrix entry, then the Smoluchowski equation with a quadratic potential is equivalent to stochastic fluctuations given by 
\begin{equation}
	\left<\delta \chi\right> = - \frac{V''(0) r}{m\,\gamma} \delta t\ \ \ ,\ \ \ 
	\left<\delta \chi^2\right> = \frac{2\,T}{m\,\gamma} \delta t\ ,
\end{equation}
which then imply the differential equations for the moments
\begin{equation}\label{eq:moment1}
	\frac{d}{dt} \left<\chi\right> = -\frac{V''(0)}{m\,\gamma} \left<\chi\right>\ ,
\end{equation}
\begin{equation}\label{eq:moment2}
	\frac{d}{dt} \left<\chi^2\right> = -\frac{2\,V''(0)}{m\,\gamma} \left<\chi^2\right>+\frac{2\,T}{m\,\gamma}\ ,
\end{equation}
The timescale of stochastic evolution can then be easily read off as
\begin{equation}\label{eq:tT}
	t\sub{T}\sim \frac{m\,\gamma}{V''(0)}\ .
\end{equation}
It is important to note that this is not the timescale over which one coarse-grains the random motion to arrive at a mean field potential for stochastic variables. This other timescale, which we call the
stochastic timescale $t\sub{stoch}$, must be shorter than the thermal timescale, $t\sub{stoch}< t\sub{T}$, and is determined from the process of averaging over microscopic dynamics. 

We next need to determine the parameters of the model. We will build this effective description of strongly coupled chaotic Matrix theory by using knowledge of the gravity dual, and of the microscopic string theory dynamics that underlies Matrix theory.   

\subsection{The light-cone Schwarzschild black hole}

We start by reviewing the dual gravity picture of the Matrix theory setup of interest -- a light-cone M-theory Schwarzschild black hole~\cite{Banks:1997cm}. The corresponding geometry is obtained by Lorentz boosting a $d$ dimensional Schwarzschild black hole in the light-cone direction with a boost factor given by $r\sub{h}/R$, where $r\sub{h}$ is the radius of the black hole horizon. 
While the horizon geometry is unchanged and the entropy or area in Planck units remains the same, the Hawking temperature is red-shifted
\begin{equation}
	T\sub{h} = \frac{R}{r\sub{h}^2}\ .
\end{equation}
The Hawking radiation flux from evaporation takes the form
\begin{equation}
	P_{\mbox{\tiny\emph{tot}}} \sim \frac{r\sub{h}^{d-1}}{r\sub{h}^{d+1}} = \frac{1}{r\sub{h}^2} 
\end{equation}
in general $d$ dimensions. The thermal timescale associated with the Hawking temperature is then
\begin{equation}\label{eq:hawkingtime}
	t\sub{h} \sim \frac{1}{T\sub{h}} \sim \frac{r\sub{h}^2}{R}\ .
\end{equation}
The entropy is related to the black hole mass $M\sub{bh}$ as usual $S\sim M\sub{bh}\, r\sub{h}$, and the evaporation process can be described by~\cite{Page:1976df,Page:1993df}
\begin{equation}
	\frac{dM\sub{bh}}{dt} = \frac{R}{r\sub{h}^3}\ .
\end{equation}
Hence, the black hole lifetime is given by 
\begin{equation}\label{eq:lifetime}
	t\sub{life} \sim t\sub{h} S\ .
\end{equation}
Beside the timescale $t\sub{h}$ and $t\sub{life}$, the shorter scrambling timescale
\begin{equation}\label{eq:scrambling}
	t\sub{scr} \sim t\sub{h} \ln S
\end{equation}
determines the timescale over which the black hole scrambles information. 
We have written all these relations in forms that can be compared to the Matrix theory stochastic model in the choice of units presented earlier. 
In our SYM choice of units, the entropy of the black hole is written as
\begin{equation}\label{eq:entropy}
	S \sim \frac{r\sub{h}^{d-1}}{\ell\sub{s}^{d-1}}\ .
\end{equation}
For a large black hole, we see that we must require
\begin{equation}\label{eq:curvature}
	r\sub{h} \gg \ell\sub{s}
\end{equation} 
leading to small curvature scales at the black hole horizon. 

The task next is to model an effective Matrix theory stochastic system that reproduces these properties of a light-cone Schwarzschild black hole.

\subsection{A conjecture for an effective model}

In a perturbative regime, Matrix theory consists of $\sim N^2$ degrees of freedom as all matrix entries participate in the dynamics. In early models of a Schwarzschild black hole in Matrix theory, the authors of~\cite{Horowitz:1997fr,Banks:1997hz,Banks:1997tn} noted however that, to reproduce the correct equation of state of a light-cone black hole, one must have the entropy proportional to $N$ at strong coupling, not $N^2$
\begin{equation}
	S \sim N\ .
\end{equation}
This implies that only $N$ of the entries in each matrix $X_i$ are to participate in the thermodynamics of the Matrix black hole; that is, most degrees of freedom must be `frozen', given that $N\gg 1$ follows from~(\ref{eq:entropy}) and~(\ref{eq:curvature}). Inspired from the works of~\cite{Horowitz:1997fr,Banks:1997hz,Banks:1997tn}, we then propose that the thermodynamics of the Matrix black hole is carried by the $N$ diagonal entries of the $X_i$ matrices. Information in the black hole would also be carried by diagonal degrees of freedom only. These entries can be sometimes interpreted as coordinates of the corresponding D0 branes underlying Matrix theory. Entropically, these order $\sim N$ degrees of freedom would like to spread to infinity -- the theory even admits flat directions for this purpose. However, perturbatively there can be an initial cost in energy in doing so from strings stretching between the D0 branes -- {\em i.e.} off-diagonal modes of the matrices. Presumably, taking strong coupling effects into account, the configuration forms a metastable ball of size $r\sub{h}$, the black hole radius, along with decay channels that implement the process of Hawking evaporation. As a diagonal matrix entry random walks its way out, a bit of the black hole evaporates away~\cite{Berkowitz:2016muc}. If $N$ diagonal degrees of freedom are to spread in a volume $r\sub{h}^{d}$, average inter-brane spacing is generically parametrically much larger with $N$ than if they are spread over an area $r\sub{h}^{d-1}$. And since inter-brane spacing is costly in energy, we can start seeing that the proper model of a Matrix black hole would involve the diagonal entries of the matrices spread on the surface of a would-be black hole horizon. Figure~\ref{fig:bh} shows a cartoon of the setup.

\begin{figure}
	\begin{center}
		\includegraphics[width=6.0in]{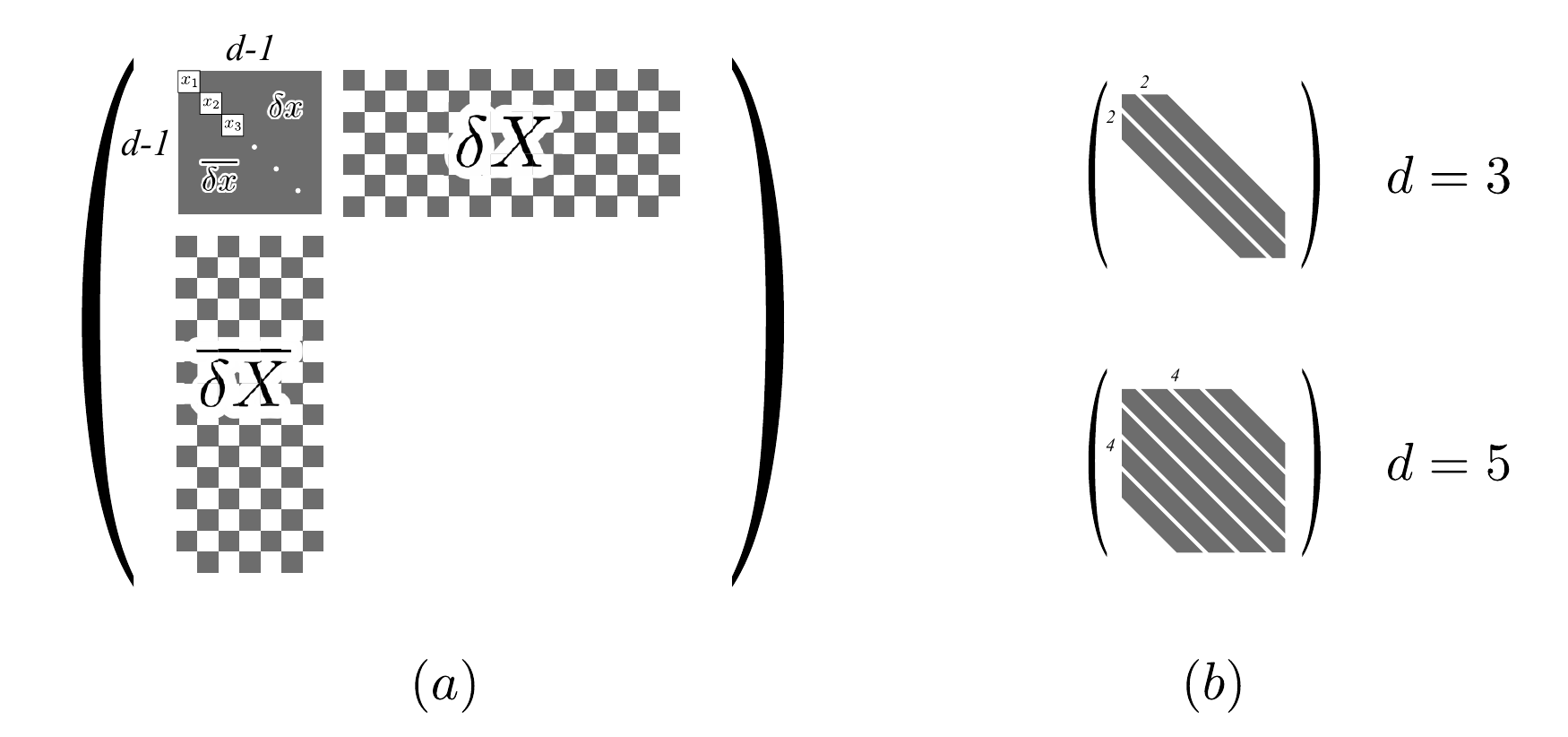}
	\end{center}
	\caption{(a) A shaded sub-block of a matrix that describes a cluster of $d-1$ D0 branes. The $\delta X$s refer to the off-diagonal entries spanning clusters; the off-diagonal entries within a cluster are in the shaded block, denoted by $\delta x$. (b) General structure of non-zero entries in the matrices for different space dimensions $d$. The $d-1$ labels refer to the number of active columns or rows in the first row or column, respectively. The shaded diagonals start within the shaded square in (a). }\label{fig:matrix}
\end{figure}
Figure~\ref{fig:matrix}(a) shows a cartoon of a matrix $X_i$, focusing on a sub-block associated with a group of `nearest-neighbor' branes\footnote{Note that the permutation symmetry requires that the additional $d^2$ off-diagonal entries in the top right and bottom left of each matrix are active as well. This is a detail in the description, in the large $N\gg d$ limit, we assume has subleading effect on the larger picture.}. Using the permutation subgroup of $U(N)$, we can always arrange to sort the matrix entries as depicted. We expect that a certain number of branes, of order $d-1$, whose coordinates appear as $x$ in the figure, would be close enough that corresponding matrix off-diagonal modes, labeled $\delta x$ in the figure, can be light. This still would not affect the $S\sim N$ requirement as the number of such modes would be independent of $N$. Branes much farther away, over a distance scale $r\sub{h}$, would be much heavier. We propose that beyond the $d\times d$ sub-block, all other off-diagonal modes would be too heavy to excite and would freeze or condense in a Bose-Einstein (BE) condensate. 
Indeed, if we look at the critical condensate temperature $T\sub{c}$, we would expect\footnote{The right hand side is the expression for the number of degrees of freedom in a Bose condensate in $d$ dimensions.} 
\begin{equation}
	N\sim N^2 \left(\frac{T\sub{h}}{T_c}\right)^{d/2}\ ,
\end{equation}
which we can quickly see to be much larger than the Hawking temperature
\begin{equation}
	T_c \sim \frac{R}{r\sub{h}^{2/d}} \gg T\sub{h}
\end{equation}
for $d\geq 2$. 
It is possible that this BE condensate describes a membrane-like configuration stretching at the black hole horizon~\cite{Horowitz:1997fr,Banks:1997hz,Banks:1997tn,Kabat:1997im,Uehara:2004vp}. In a coarse-grained effective language, we would set these heavy off-diagonal modes, the $\delta X$s in the figure, to zero. Interestingly, fuzzy spheres of various dimensions in Matrix theory have been shown to necessitate the activation of more off-diagonal modes that spread away from the diagonal~\cite{Castelino:1997rv,Ramgoolam:2001zx}. For example, a 2-sphere ($d=3$) is realized through $SU(2)$ representations, which activate $3$ diagonal lines along the matrix diagonals; and a 4-sphere ($d=5$) activates $5$ diagonal lines. Our model then fits well with this pattern. Figure~\ref{fig:matrix}(b) shows the general scheme. 

The diagonal entries {\em within} the $d\times d$ sub-block of matrices would be spread out from
each other at a distance that is around the Planck scale and might naturally involve marginal bound state physics. In M-theory language, this would correspond to supergravity excitations carrying $\sim d$ units of light-cone momentum. These marginal bound states are conjectured to exist in Matrix theory and are a necessary ingredient for the dictionary between Matrix theory and M-theory~\cite{Banks:1996vh}. The off-diagonal modes $\delta x$ in these sub-blocks would remain relatively light and participate in making the physics of these clusters non-local, at around the Planck scale. They would correspond to strings joining nearest neighbor branes, and henceforth we refer to the $\delta x$s as `{\em off-diagonal nearest neighbor modes}'\footnote{Our treatment explicitly picks out a `frame' or gauge where the diagonal and off-diagonal matrix entries have very different physical roles. We expect that this setup corresponds to a description of the Matrix black hole from the perspective of the outside observer. $U(N)$ gauge transformations would naturally change the perspective, while mixing the roles of diagonal and off-diagonal entries. More on this in the Discussion section.}.

Our stochastic model would then involve writing an effective theory of all the modes that remain active -- diagonals $x$ and nearest neighbor modes $\delta x$ -- while integrating out all other $\delta X$ modes. We need to provide two separate stochastic treatments, one for the $x$ modes on the diagonal, and another for the off-diagonal nearest neighbor modes $\delta x$. The first would describe the coarse-grained thermal state of the black hole; the second would describe 
finer cluster physics within each matrix sub-block. We will next demonstrate how these two sectors effectively decouple and can reliably be treated through stochastic methods due to a hierarchy in the relevant timescales.

In the Matrix theory scaling regime time scales as $g\sub{s}/\ell\sub{s}$; this allows us to measure timescale through the effective Yang-Mills coupling $g\sub{eff}(\tau)^2$ defined as
\begin{equation}
	\frac{g\sub{s}}{\ell\sub{s}} t = \frac{g\sub{s}^{1/3}}{\ell\sub{s}} (g\sub{s}^{2/3}\,t)  = (g\sub{YM}^2)^{1/3} \tau \equiv (g\sub{eff}(\tau)^2)^{1/3}\ ,
\end{equation}
which remains finite in the scaling regime. Hence, larger effective coupling corresponds to longer times since $0+1$ SYM is super-renormalizable. In this language, the first timescale $t\sub{h}$ from~(\ref{eq:hawkingtime}) arises from the thermodynamics of the diagonal modes, of order $N$ in number; this gives
\begin{equation}
	\frac{g\sub{s}}{\ell\sub{s}} t\sub{h} = (g\sub{eff}(\tau\sub{h})^2)^{1/3} \sim \left(\frac{r\sub{h}}{\ell\sub{s}}\right)^2\gg 1\ .
\end{equation}
The scrambling timescale $t\sub{scr}$ of~(\ref{eq:scrambling}) is then given by
\begin{equation}
	\frac{g\sub{s}}{\ell\sub{s}} t\sub{scr} =(g\sub{eff}(\tau\sub{scr})^2)^{1/3} \sim \ln N\,\left(\frac{r\sub{h}}{\ell\sub{s}}\right)^2\gg 1\ .
\end{equation}
The lifetime of the configuration $t\sub{life}$ from~(\ref{eq:lifetime}) should correspond to 
\begin{equation}
	\frac{g\sub{s}}{\ell\sub{s}} t\sub{life} = (g\sub{eff}(\tau\sub{life})^2)^{1/3} \sim \left(\frac{r\sub{h}}{\ell\sub{s}}\right)^2 N \gg 1\ .
\end{equation}
These statements follow from the expected black hole physics on the dual side of the correspondence. Note that all three timescales correspond to regimes where the Matrix theory SYM is strongly coupled.

On the SYM side, perturbatively, we know that off-diagonal modes have dynamics given by\footnote{The total energy receives an important contribution from fermionic zero modes which will be taken into account when developing the mean field potential. At this stage, we use the bosonic sector only to simply identify relevant dynamical scales. Note also that, at finite temperature, supersymmetry would be broken.}
\begin{equation}
	E\sim \frac{1}{R} \delta\dot{x}^2+\frac{R}{\ell\sub{s}^6} \Delta r^2 \delta x^2\ ,
\end{equation}
where $\Delta r$ is the distance between the corresponding diagonal entries; this gives a frequency of
\begin{equation}\label{eq:deltar}
	\omega_{\delta x} \sim \frac{R}{\ell\sub{s}^2}\frac{\Delta r}{\ell\sub{s}} \ .
\end{equation}
We can then easily see that if $\Delta r\sim \ell\sub{s}$ for nearest neighbor off-diagonal modes, 
$\delta x$ modes can be treated as heavy and can hence be integrated out over time scales 
\begin{equation}
	\omega_{\delta x} t > 1\Rightarrow t>t\sub{o}\ \ \ \mbox{with}\ \ \ \frac{g\sub{s}}{\ell\sub{s}}t\sub{o} = 1 \Rightarrow (g\sub{eff}(\tau)^2) > 1\ .
\end{equation}
This is the strong coupling transition point for the SYM, a regime that we typically associate with emergence of geometry on the dual M-theory side. The relevant strong coupling benchmark is given by  $g\sub{eff}(\tau)^2\sim 1$, instead of the one using the 't Hooft effective coupling $g\sub{eff}(\tau)^2 N \sim 1$, because the dynamics in question is that of individual partons in the black hole soup, as opposed to the interaction of the black hole as a whole. More on the interplay between these two couplings and the emergence of a valid geometrical description can be found in the Discussion section. 

Next, looking at off-diagonal modes $\delta X$ that straddle diagonal modes separated by a large distance of order $\Delta r\sim r\sub{h}$, we see from~(\ref{eq:deltar}) that these can be integrated out for timescales
\begin{equation}
	\omega_{\delta X} t\gg 1\Rightarrow t\gg t\sub{stoch}\ \ \ \mbox{with}\ \ \ \frac{g\sub{s}}{\ell\sub{s}} t\sub{stoch} = (g\sub{eff}^2(\tau\sub{stoch}))^{1/3} = \frac{\ell\sub{s}}{r\sub{h}} \Rightarrow (g\sub{eff}(\tau)^2)^{1/3} \gg \frac{\ell\sub{s}}{r\sub{h}}\ .
\end{equation}
This is the shortest of the timescales and determines the regime where a stochastic treatment is valid: it corresponds to timescales where integrating out the $\delta X$'s leads to a stochastic mean field potential for the diagonal modes. Note also that, for $r\sub{h}\gg \ell\sub{s}$, part of this regime overlaps with weak coupling in the Matrix SYM.

\begin{figure}
	\begin{center}
		\includegraphics[width=6in]{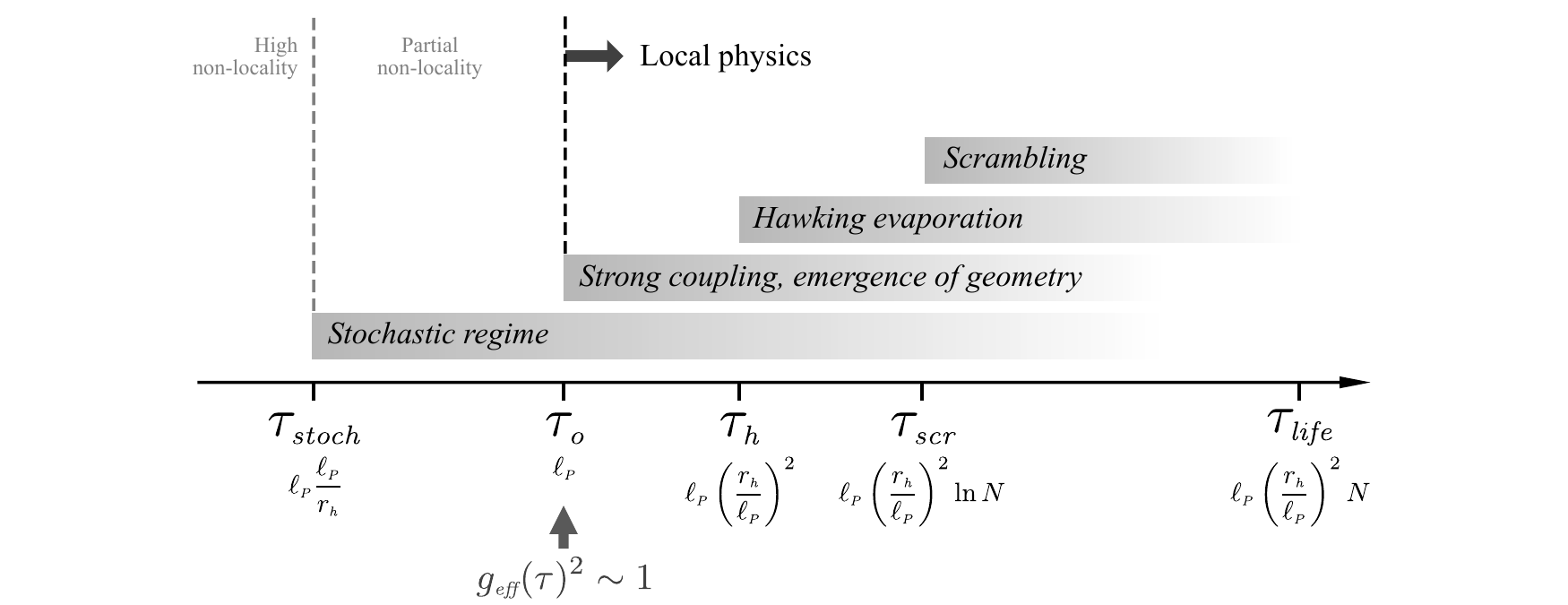}
	\end{center}
	\caption{The hierarchy of timescales for event horizon dynamics. Timescales $t<t\sub{o}$ are 
associated with non-local physics within D0 brane clusters, but timescales $t>t\sub{stoch}$ allow a local description for coarser inter-cluster dynamics.}\label{fig:timescales}
\end{figure}
Figure~\ref{fig:timescales} summarizes the various timescales and clarifies the range of validity for the effective model that we propose. The stochastic formalism with a mean field potential for the diagonal modes  requires coarse graining over time scales longer than $t\sub{stoch}$.  For $t>t\sub{stoch}$, $\delta X$'s are frozen in a BE condensate. We can then incorporate the effect of the $\delta X$'s into a mean field potential for the modes on the diagonal. The nearest neighbor off-diagonal modes, the $\delta x$'s, cannot be integrated out at these timescales. We leave them part of the degrees of freedom participating in the physics of 
cluster formation. For timescales $t>t\sub{o}$, the nearest neighbor modes are heavy as well and are associated with  high frequency dynamics that can be coarse grained and described through a stochastic treatment. However, the $\delta X$ modes will always have a much higher frequency (for $r\sub{h}\gg\ell\sub{s}$) and hence will still determine the mean field potential for the diagonal modes. Finally, thermal timescales, $t\sub{h}$, $t\sub{scr}$, and $t\sub{life}$ are all much longer and live well within the regime of a stochastic treatment that coarse grains physics faster than $t\sub{stoch}$. 

We then list in one place the set of observations underlying our model:

\begin{itemize}
	\item We have a stochastic effective description for diagonal modes for $t>t\sub{stoch}$, or\\ $(g\sub{eff}(\tau)^2)^{1/3} \gg \frac{\ell\sub{s}}{r\sub{h}}$. We integrate out the off-diagonal modes that straddle widely separated modes on the diagonal.
	\item Strong coupling corresponds to timescales $t>t\sub{o}$, or $(g\sub{eff}(\tau)^2)^{1/3} \gg 1$. In this regime, all off-diagonal modes are heavy, but the effect of nearest neighbor off-diagonal modes on diagonal modes is sub-leading. We associate emergence of geometry on the dual M-theory side with the onset of strong coupling in Matrix theory~\cite{Bigatti:1997jy,Taylor:2001vb}. At timescales $t\sub{stoch} < t \lesssim t\sub{o}$, 
we might be able to write a stochastic effective description of D0 brane cluster dynamics. We expect that at around $t\sim t\sub{o}$, the degrees of freedom of Matrix theory organize in clusters of about $d$ nearest neighbor branes moving in the larger thermal soup.
	\item Hawking evaporation physics sets in at $t\gtrsim t\sub{h}$, or $(g\sub{eff}(\tau)^2)^{1/3} \sim \left(\frac{r\sub{h}}{\ell\sub{s}}\right)^2\gg 1$, well within the regime of validity of the stochastic treatment. 
\end{itemize}

It is useful to write some of these timescales in M-theory Planck units. Using~(\ref{eq:rescaled}), and the fact that light-cone time is boosted by a factor of $\ell\sub{P}/R$, we find
\begin{equation}
	\tau\sub{o} = \frac{\ell\sub{P}}{R} \ell\sub{P} \rightarrow \ell\sub{P}\ ,
\end{equation}
\begin{equation}
	\tau\sub{stoch} = \frac{\ell\sub{P}}{R} \ell\sub{P} \frac{\ell\sub{P}}{r\sub{h}} \rightarrow \ell\sub{P} \frac{\ell\sub{P}}{r\sub{h}} \ll \ell\sub{P}\ ,
\end{equation}
and
\begin{equation}
	\tau\sub{h} = \frac{\ell\sub{P}}{R} \ell\sub{P} \left(\frac{r\sub{h}}{\ell\sub{P}}\right)^2 \rightarrow \ell\sub{P} \left(\frac{r\sub{h}}{\ell\sub{P}}\right)^2 \gg \ell\sub{P}\ .
\end{equation}
Hence we see that $\tau\sub{o}$ correspond to Planck scale time in M-theory language. As we shall see, all this means that the chaotic microscopic dynamics that underlies black hole horizon physics is associated with a characteristic timescale that is given by the Planck scale. A well-defined notion of spacetime geometry necessitates coarse graining over longer timescales.

Our next task is to develop the stochastic effective descriptions of diagonal and nearest neighbor off-diagonal modes -- the first describing black hole thermodynamics and evaporation, the second giving us a crude peak into brane cluster/bound state dynamics.

\subsection{Modes on the diagonal}

In this section, we propose a mean field stochastic potential for diagonal modes, valid over timescales $t>t\sub{stoch}$. Using spherical coordinates, we posit
\begin{equation}\label{eq:potential}
	V(r) = - V_0 \left(\frac{r^2}{r_0^2}-1\right)^2 \theta(r_0-r)\ ,
\end{equation}
writing $r^2=x_i^2$, where $x_i$ is any diagonal mode of $X_i$. The potential is parametrized by two scales, $r_0$ and $V_0$, and we need to determine these two parameters by comparing the resulting dynamics to that of a light-cone black hole. Note also that we have incorporated quantum effects that we know would arise from the fermionic sector of Matrix theory: the $\theta(r_0-r)$ flattens the potential so as to model the expected flattening of the potenial from supersymmetry-based cancellations of zero mode energies\footnote{The potential is not strictly flat but comes with an $1/r^{d-2}$ fall-off at one loop order. For the purposes of the approximate stochastic description, we treat this as flat since no aspect of the model explores the region far away from the black hole.}. 

We start by noting that the only scale near the horizon of the Schwarzschild black hole is given by $r\sub{h}$ \footnote{This might prejudice the discussion in favor of black hole complementarity~\cite{Susskind:1993if}-\cite{Susskind:2012rm} as opposed to a firewall scenario at the horizon~\cite{Almheiri:2012rt}-\cite{Harlow:2013tf}. Nevertheless, we still need to map onto geometry on the dual M-theory side. We have tried to develop a model with an additional scale in the mean field potential set at the Planck scale near the horizon, and it seems that this does not lead to a picture that is consistent with Hawking evaporation. While we cannot rule out the possibility of finding  an alternate model that includes the Planck scale -- as we have not explored all possibilities -- we note however that the simple model given in the text works very well without the need of a Planck scale at the horizon.}. We then start by setting
\begin{equation}
	r_0 = r\sub{h}
\end{equation}
fixing the size of the stochastic diagonal fluctuations to within the would-be horizon size. 
The temperature of the soup should naturally be the Hawking temperature in the light-cone frame
\begin{equation}
	T=T\sub{h} \simeq \frac{R}{r\sub{h}^2}\ .
\end{equation}
The mass of a stochastic particle should be set to the mass of a D0 brane
\begin{equation}
	m = \frac{1}{R}\ .
\end{equation}
This leaves us with determining the damping parameter $\gamma$ and the potential scale $V_0$. We start by looking at evaporation flux from the thermal soup. Following~\cite{weiss}, we arrange for a steady state scenario for the probability distribution given by
\begin{equation}
	p = C\, f(u) \exp\left[-\frac{1}{T} \left(\frac{1}{2}m\,v^2+V(r)\right)\right]\ ,
\end{equation}
where $u=r-r_0$ and $C$ is a normalization constant to be determined. We need to find $f(u)$ given the boundary conditions
\begin{equation}
	f(-r_0) = 1\ \ \ \mbox{and}\ \ \ f(0)\simeq 0\ ,
\end{equation}
where the first one follows from matching with the equilibrium configuration at $r=0$, while the second one amounts to absorbing the evaporation flux at $r=r_0$, corresponding to evaporation to infinity. The Fokker-Planck equation at strong damping then leads to  
\begin{equation}
	\kappa u\,f'(u)+f''(u) = 0\ ,
\end{equation}
where
\begin{equation}
	\kappa = -\frac{V''(r_0)}{T} =  \frac{8\,V_0}{T\,r_0^2} > 0
\end{equation}
for the mean field potential at hand. The solution is given by the error function
\begin{equation}
	f(u) = \frac{\mbox{erf}((r-r_0) \sqrt{\kappa/2})}{\mbox{erf}(-r_0 \sqrt{\kappa/2})}\ .
\end{equation}
Integrating over the velocities we have
\begin{equation}
	p = C\, f(u) e^{-V(r)/T} \left(\frac{2\,\pi\,T}{m}\right)^{d/2}\ ,
\end{equation}
which then leads to the current
\begin{equation}
	j_r = -C\, \frac{T}{m\,\gamma} \left(\frac{2\pi\, T}{m}\right)^{d/2} \sqrt{\frac{2\,\kappa}{\pi}} e^{-V(r)/T} \frac{e^{-\frac{\kappa}{2} (r-r_0)^2}}{\mbox{erf}(-r_0 \sqrt{\kappa/2})}\ .
\end{equation}
We will see below that
\begin{equation}
	r_0 \sqrt{\frac{\kappa}{2}} \simeq \sqrt{\frac{V_0}{T}} \simeq 1\ ,
\end{equation}
when we find that $V_0 \sim T\sub{h}$. We then note that 
\begin{equation}
	\mbox{erf}(-r_0 \sqrt{\kappa/2}) \simeq -1\ .
\end{equation}
For $\mbox{erf}(-x)$, the function near $x \gtrsim 1$ is very well approximated by $-1$ with corrections suppressed exponentially as $e^{-x^2}/x$.
We determine the normalization factor $C$ using
\begin{equation}
1 = \int \mbox{d}^d{\bm{r}}\, p(r,t)\ .
\end{equation}
For this, we write
\begin{equation}
	f(r-r_0) \simeq 1 + \frac{2}{\sqrt{\pi}} \sqrt{\frac{\kappa}{2}} r\,e^{-r_0 \sqrt{\kappa/2}}
\end{equation}
near $r\simeq 0$, and
\begin{equation}
	f(r-r_0) \simeq \frac{2}{\sqrt{\pi}}\sqrt{\frac{\kappa}{2}} (r-r_0)
\end{equation}
near $r\simeq r_0$. 
We then get
\begin{equation}
	1\simeq C\, \frac{T^d}{m^{d/2}} e^{V_0/T} \frac{r_0^{d}}{V_0^{d/2}}
\end{equation}
up to a numerical factor. The probability current near $r_0$ takes the form
\begin{equation}
	j(r_0) \simeq C\, \frac{T}{m\,\gamma}\left(\frac{2\pi\,T}{m}\right)^{d/2} e^{-V(r_0)/T}\ ,
\end{equation}
which then leads to the evaporation flux
\begin{equation}
	F \simeq j(r_0) r_0^{d-1} \simeq \frac{T^{-(d-1)/2}V_0^{(d+1)/2}}{m\,r_0^2\gamma}\ ,
\end{equation}
which we can then match with Hawking evaporation at temperature $T\sub{h}$
\footnote{
If we want to include the kinetic energy of the evaporated bit, we would get 
\begin{equation}
	F \simeq \frac{T^{-(d-1)/2}V_0^{(d+1)/2}}{m\,r_0^2\gamma} e^{-(\omega+V_0)/T}\ ,
\end{equation}
with $\omega$ being the kinetic energy, giving the standard black body spectrum.
}
\begin{equation}
	F=F\sub{h}\simeq \frac{R}{r_0^2}\ .
\end{equation}
This gives one of the two conditions we need to determine $\gamma$ and $V_0$. The other condition comes from the well-known one-loop effective potential of a probe D0 brane in the background of $N$ D0 branes.
Using M-theory Planck units, we have~\cite{Kabat:1997im}
\begin{equation}
	V\simeq \frac{N\, \ell\sub{P}^{d-1} v^4}{R^3 r^{d-3}}\ ,
\end{equation}
where $v$ is the relative velocity of two partons at a separation $r\sim r\sub{h}$. While this is a perturbative result in the Matrix SYM, it is know to lead to an exact match with the dual M-theory scenario~\cite{Kabat:1997im} implying that it is valid at strong coupling as well\footnote{There have been suggestions that a non-remormalization theorem perhaps underlies this finding~\cite{Bigatti:1997jy}.}. Remembering that the black hole entropy is given by
\begin{equation}
	S\simeq \frac{r\sub{h}^{d-1}}{\ell\sub{P}^{d-1}} \sim N
\end{equation}
in Planck units, and saturating the Heisenberg uncertainty bound for each parton~\cite{Horowitz:1997fr,Banks:1997hz,Banks:1997tn}
\begin{equation}
	v\sim \frac{R}{r\sub{h}}\ ,
\end{equation}
we get the scale of the potential energy at the size of the horizon
\begin{equation}
	E\simeq \frac{R}{r\sub{h}^2}\ .
\end{equation}
Rescaling to SYM units using~(\ref{eq:rescaled}) gives the same relation ($r\sub{h}\rightarrow r\sub{h}\sqrt{R}$, $E\rightarrow E/R$). We then naturally identify this energy scale with the depth of the mean field potential
\begin{equation}\label{eq:Vdiagonal}
	V(0)\simeq E \Rightarrow V_0 = \frac{R}{r\sub{h}^2} = T\sub{h}\ .
\end{equation}
Finally, from $F=F\sub{h}$, we then get
\begin{equation}
	\gamma \simeq \frac{R}{r\sub{h}^2}\ .
\end{equation}
The latter relation implies that
\begin{equation}
	m\,\gamma \simeq \frac{1}{r_0^2}\ ,
\end{equation}
which corresponds to a borderline strong damping regime~(\ref{eq:strongdamping}) -- needed for consistency with RMT.

We can now look at the quantum and thermal vacuum expectation values of a mode $x$ on the diagonal, given by
\begin{equation}
	\left<x^2\right>\sub{th} \sim \frac{T}{V''(0)}\ \ \ ,\ \ \ 
	\left<x^2\right>\sub{qu} \sim \sqrt{\frac{R}{V''(0)}}\ .
\end{equation}
For the given potential and parameters, we have
\begin{equation}
	\left<x^2\right>\sub{th} \simeq \left<x^2\right>\sub{qu} \sim r\sub{h}^2\ ,
\end{equation}
leading to borderline thermal regime,
which implies that the diagonal modes are barely excited above the ground state.
We also note that odd moments vanish at equilibrium, so that
\begin{equation}
	\left<x\right>=0\ .
\end{equation}

We then have succeeded in developing a stochastic model for diagonal mode dynamics that matches with Hawking evaporation. As a result, a consistency check shows that this stochastic evolution has characteristic timescale given by~(\ref{eq:tT})
\begin{equation}
	t\sub{T} \sim \frac{m\,\gamma\,r\sub{h}^2}{V_0} = 
		 \frac{r\sub{h}^2}{R} = t\sub{h}
\end{equation}
as required.

\subsection{Off-diagonal nearest neighbor modes}

At timescales $t\sim t\sub{o}$, where Matrix theory enters the strongly coupled realm, we have the possibility to describe clusters of $d$ nearest neighbor branes through stochastic means. The clusters are marginally held together and we expect this dynamics to be a delicate one, given their natural overlap with the physics D0 brane marginal bound state formation. Nevertheless, we will use the methods of stochastic dynamics to try to describe the problem, bearing in mind that we aim only to identify scaling relations of what is most likely a very subtle cluster formation process. We model the potential for the  nearest neighbor off-diagonal modes $V_{\delta x}$ with a simple quadratic confining form, and the only relevant scale is the curvature $V''(0)$. For nearest neighbor diagonals, we expect an inter-brane separation of $\Delta r\sim \ell\sub{s}$, leading to a perturbative potential for the corresponding off-diagonal modes given by
\begin{equation}\label{eq:Vdeltax}
	V''_{\delta x}(0) \sim \frac{R}{\ell\sub{s}^6} \Delta r^2 \sim \frac{R}{\ell\sub{s}^4} = \frac{g\sub{s}}{\ell\sub{s}^3} = g\sub{YM}^2\ .
\end{equation}
This is a perturbative result but we extend it to $t\lesssim t\sub{o}$ as a scaling relation. 
The thermal and quantum vacuum expectation values are 
\begin{equation}
	\left<\delta x^2\right>\sub{th} \sim \frac{T}{V''(0)} = \frac{T}{g\sub{YM}^2}\ \ \ ,\ \ \ 
	\left<\delta x^2\right>\sub{qu} \sim \sqrt{\frac{R}{V''(0)}} = \sqrt{\frac{R}{g\sub{YM}^2}}\ ,
\end{equation}
where in the thermal expression, we want to think of $T$ as a scale for kinetic energy within the bound system.
We would expect ground state physics, implying
\begin{equation}
	\left<\delta x^2\right>\sub{th}\sim \left<\delta x^2\right>\sub{qu} \Rightarrow \frac{T^2}{g\sub{YM}^2} \sim R\ ,
\end{equation}
which identifies
\begin{equation}
	T_{\delta x} \sim \frac{R}{\ell\sub{s}^2}
\end{equation}
as the expected scale for kinetic energy in the cluster. The mass parameter would still be given by 
\begin{equation}
	m_{\delta x}=\frac{1}{R}\ .
\end{equation}
Finally, we propose that the strong damping bound needed by RMT should be valid, and at worst saturated
\begin{equation}
	(m\,\gamma)^2 \sim m\,V''(0) \Rightarrow m\,\gamma \sim \frac{1}{\ell\sub{s}^2}\ ,
\end{equation}
identifying the damping parameter $\gamma$ for cluster dynamics. 
As a sanity check, we can verify that the associated characteristic timescale for the stochastic dynamics is 
\begin{equation}
	\mbox{Timescale} \sim \frac{m\,\gamma}{V''(0)} \sim \frac{\ell\sub{s}}{g\sub{s}} = t\sub{o}\ ,
\end{equation}
which again matches well with our expectations that the relevant dynamics is at the onset of 
strong coupling in the SYM theory. Finally, the expected size of the cluster becomes
\begin{equation}
	\mbox{Size}^2 \sim \frac{T}{V''(0)} \sim \ell\sub{s}^2\ ,
\end{equation}
which also syncs well with our expectation that one thermal parton is to occupy one Planck area at the black hole horizon
\footnote{
Note that in M-theory Planck units, this translates to $\mbox{Size}\sim \ell\sub{P}$
as expected, given that $X\rightarrow X/\sqrt{R}$.
}.

\section{Quantum information}

In this section, we want to describe how information evolves in the stochastic model we developed above. For this purpose, we need to look more closely at the fermionic degrees of freedom of the $\Psi$ matrix in~(\ref{eq:sym}). It is known that these correspond to the polarizations of the light-cone M-theory supergravity multiplet -- the graviton, the gravitino, and the 3-form gauge field~\cite{Banks:1996vh}. That is, in the low energy regime, we can think of an entry on the diagonal in the $X_i$'s as the coordinate of a supergravity particle whose flavor and polarization state is determined by the corresponding entry in the $\Psi$ matrix. We can expect that information in an M-theory black hole can be encoded in the polarization states of a thermal soup of supergravity excitations. We would then want to study the time evolution of the $\Psi$ matrix within the effective model we have developed. Note that the quantum contribution from the fermionic modes in their ground state has already been taken into account 
in the shape of the mean field potential for the diagonal bosonic modes. 

In the spirit of RMT, the equilibrium dynamics of the fermionic and bosonic matrix entries are treated as statistically uncorrelated. This justifies working with the bosonic sector by itself as we have done so: it is assumed that a corresponding thermal state is also set up in the fermionic sector as the two sectors are in thermal equilibrium. Our goal now is to track how information encoded in the polarization states evolves when this equilibrium configuration is slightly perturbed. We could for example consider one particularly interesting scenario, the emission of a supergravity particle from the stochastic soup, as a matrix entry of $X_i$ ventures off to large distances. We would choose a particular matrix configuration that can describe this situation, and analyze the evolution of the corresponding bit of quantum information in $\Psi$. 

\subsection{Qubit dynamics and M-theory polarizations}

We start by considering a $d=3$ matrix configuration that looks like\footnote{The $\Delta X$s in this expression are set to zero to leading order in the computation as they are fast modes frozen in the vacuum and their effect is already incorporated in the mean-field potential. The expectation values $\left<\delta X\right>$ in the vacuum scales inversely with the large frequency.}
\begin{equation}\label{eq:matrices}
	X = \left(
	\begin{array}{ccc}
		X\sub{bh} & \delta x\sub{bh} & 0 \\
		\overline{\delta x}\sub{bh} & x\sub{bh} & \delta x \\
		0 & \overline{\delta x} & x
	\end{array}
	\right) \ \ \ ,\ \ \ 
	\Psi = \left(
	\begin{array}{ccc}
		\Psi\sub{bh} & \delta \psi\sub{bh} & 0 \\
		\overline{\delta \psi}\sub{bh} & \psi\sub{bh} & \delta \psi \\
		0 & \overline{\delta \psi} & \psi
	\end{array}
	\right)\ ,
\end{equation}
where $X\sub{bh}$ and $\Psi\sub{bh}$ are a $(N-2)\times (N-2)$ sub-blocks representing part of the black hole, and the remaining $x\sub{bh}$/$\psi\sub{bh}$ and $x$/$\psi$ represent $1\times 1$ entries that are bits of the black hole that will participate in an emission process. The particle with coordinate $x$ and polarization state $\psi$ has perhaps ventured outside the black hole via ergodic motion. The $\delta x$ mode is a nearest neighbor off-diagonal, implying that $x\sub{bh}$ and $x$ are part of a cluster. The rest of the matrix entries start off in an equilibrium state at temperature $T\sub{h}$. Note that $\delta x\sub{bh}$ and $\delta \psi\sub{bh}$ are $N-2$ component vectors. The fermionic part of the Matrix theory action is given by~(\ref{eq:sym})
\begin{equation}\label{eq:stochaction}
	S\sub{ferm}[X,\Psi]=\int dt\, \frac{1}{2} \Psi \dot{\Psi} + \frac{R}{2\,\lambda^{3/2}} \Psi \Gamma^i \left[X_i, \Psi\right]\ .
\end{equation}
Quantizing the fermionic matrix entries, we have
\begin{equation}
	\left\{\Psi_{ab\, \alpha}, \Psi^\dagger_{ab\, \beta}\right\} = 2\, \delta_{\alpha\beta}\ ,
\end{equation}
where $\alpha$ and $\beta$ are 10 dimensional spinor indices, $\alpha, \beta = 1,\ldots, 16$, remembering that the matrix entries $\Psi_{ab}$ are Majorana-Weyl in $10$ spacetime dimensions. Applying this quantization to the matrix configuration~(\ref{eq:matrices}), we get for the off-diagonal modes 
\begin{equation}
	\left\{\delta \psi_\alpha, \delta\overline{\psi}_\beta\right\} = 2\,\delta_{\alpha\beta}\ ,
\end{equation}
while the diagonal entries lead to a Clifford algebra
\begin{equation}
	\left\{\psi_\alpha, \psi_\beta\right\} = 2\,\delta_{\alpha\beta}\ .
\end{equation}
The latter means that we can introduce new raising/lowering spinors on the diagonal by
\begin{equation}\label{eq:diagqubits}
	\psi^\pm_\alpha = \frac{1}{2} \left(\psi_\alpha \pm i\, \psi_{\alpha+8}\right)
\end{equation}
where we now restrict $\alpha = 1, \ldots, 8$. We then have
\begin{equation}
	\left\{\psi^+_\alpha,\psi^-_\beta\right\} = \delta_{\alpha\beta}
\end{equation}
as needed. In general, the fermionic sector then consists of $8\,N\,(N-1)$ qubits from off-diagonal modes and $8\,N$ qubits from the diagonal modes for a total of $8\,N^2$ qubits corresponding to $2^8=256$ polarization states of the M-theory supergravity multiplet -- one for each of the $N^2$ matrix degrees of freedom. 

Using~(\ref{eq:matrices}), we can then expand the action~(\ref{eq:stochaction})
treating all matrix entries as stochastic variables. Furthermore, given spherical symmetry, we expect all spatial directions to be statistically equivalent so that we can write $x_i \rightarrow x$ for all $i$. We get the action 
\begin{eqnarray}\label{eq:qubitaction}
	&&S\sub{ferm} = \frac{2\,\sqrt{d}}{(2\pi)^{3/2}}\,\frac{R}{\ell\sub{s}^3}\, \left[\left((x-x\sub{bh})\, \overline{\delta\psi}\, \Gamma \delta\psi
	+\delta x\, \overline{\delta\psi}\, \Gamma (\psi-\psi\sub{bh})
	- \delta \psi\, \Gamma (\psi-\psi\sub{bh})\, \overline{\delta x}\right)\right. \nonumber \\
	&&+	\left. \left(\overline{\delta \Psi}\sub{bh}\, \Gamma (X\sub{bh}-x\sub{bh})\delta \Psi\sub{bh}
	-\overline{\delta\Psi}\sub{bh}\,\Gamma (\Psi\sub{bh}-\psi\sub{bh})\,\delta X\sub{bh}
	-\overline{\delta X}\sub{bh}\,(\Psi\sub{bh}-\psi\sub{bh})\, \Gamma \delta\Psi\sub{bh}\right)\right]
\end{eqnarray}
where we define
\begin{equation}
	\Gamma \equiv \frac{1}{\sqrt{d}} \sum_i \Gamma_i\ .
\end{equation}
Throughout, we use a symmetric representation for the $\Gamma_i$s. 
Note that $\Gamma^2=1$ and $\mbox{Tr}\, \Gamma =0$ so that the eigenvalues of $\Gamma$ are $\pm 1$.
We will then choose the convenient representation where
\begin{equation}\label{eq:Gamma}
	\Gamma = \left(
		\begin{array}{cc}
			1_{8\times8}	&	0_{8\times8} \\
			0_{8\times8}	&	-1_{8\times8}
		\end{array}
	\right)\ .
\end{equation}

Taking the thermal vacuum expectation value of~(\ref{eq:qubitaction}), we see that the thermal average of the action $\left<S\sub{ferm}\right>$ vanishes at equilibrium given that we know
\begin{equation}
	\left< x \right> = \left< x\sub{bh} \right> = \left< \delta x \right> = \left< X\sub{bh} \right> = \left< \delta X\sub{bh} \right> = 0\ .
\end{equation}
This is simply the statement that, once equilibrium is achieved, we have two separate systems -- a bosonic and a fermionic one -- that can be treated as two thermal components in equilibrium at the same temperature. The interesting physics arises when we consider a perturbed configuration, for example one corresponding to $x-x\sub{bh}$ being momentarily large -- describing the process of evaporation of a bit of the Matrix black hole. The subsequent relaxation process would be driven by  the couplings in~(\ref{eq:qubitaction}) between bosonic modes and qubits. We can analyze this physical setup by looking at the stochastic effective action of the qubits provided we arrange proper boundary conditions where $x$ and $\delta x$ are initially perturbed away from equilibrium. In the next section, we develop this method of tracking qubit information evolution.

\subsection{Qubit action}

We expect that a small perturbation should not affect the whole system appreciably on short enough timescales. This means that if we were to perturb $x$ and $\delta x$ in~(\ref{eq:matrices}) off-equilibrium, $X\sub{bh}$ and $\delta X\sub{bh}$ (as well as $\Psi\sub{bh}$ and $\delta \Psi\sub{bh}$) would remain in equilibrium as long as $N\gg 1$.  Using techniques from~\cite{janssen}, given a stochastic variable $\chi$ coupling to other degrees of freedom $F(t)$ via $S=\int dt\, \chi\,F$, we can write an action 
\begin{equation}\label{eq:Seffstoch}
	i\,S\sub{q} = \ln \int \mathcal{D}\chi\, e^{-S\sub{stoch}[\chi] + i\, \int dt\,\chi F(t)}\ ,
\end{equation}
where $\chi$ would be $x$ or $\delta x$ from earlier, and where 
the Stochastic action is
\begin{equation}
	S\sub{stoch}[\chi] = \int dt \left[\frac{i\,m\,\gamma}{4\,T} \left(\dot{\chi} + \frac{V'(\chi)}{m\,\gamma}\right)^2
		-\frac{i}{2\,m\,\gamma} V''(\chi)
	\right]
\end{equation}
$T$ is the temperature to which the perturbed $\chi$ relaxes to, and the path integration involves boundary conditions corresponding to the quenching process of interest. The potential $V$, the damping parameter $\gamma$, and the mass $m$ are all determined from our previous discussion in Section~\ref{sec:first}. $F(t)$ can be obtained from~(\ref{eq:qubitaction}) and is bilinear in the qubit variables. It can easily be shown that the Smoluchowski equation for $\chi$ given by~(\ref{eq:smol}) follows from $S\sub{stoch}$~\cite{janssen}. 

To evaluate the path integral, we start with the classical equations of motion
\begin{equation}
	\frac{i\,m\,\gamma}{2}\frac{1}{T} \ddot{\chi} - \frac{i}{2} \frac{m\,\gamma}{T} \Omega^2 \chi = F\ ,
\end{equation}
where
\begin{equation}\label{eq:Omega}
	\Omega \equiv \frac{V''(0)}{m\,\gamma}\ .
\end{equation}
If $\chi$ represents a radial coordinate $\sqrt{x_i^2}$ in a spherically symmetric setup as given by~(\ref{eq:potential}) we get instead
\begin{equation}
	\Omega^2 = \frac{V_0}{(m\,\gamma)^2\,r_0^2} \left(16\,\frac{V_0}{r_0^2}+8\,(d+2)\, \frac{T}{r_0^2}\right)\ .
\end{equation}
Since $V_0 \sim T$ for any of the bosonic perturbations of interest, $\Omega$ has then the same scale irrespective of symmetry. We solve the sourceless classical equation
\begin{equation}
	\ddot{\chi}-\Omega^2 \chi = 0\ ,
\end{equation}
and we easily find
\begin{equation}\label{eq:chicl}
	\chi\sub{cl}(t) = \frac{1}{\sinh \Omega (t_i-t_f)}\left[ \chi_i \sinh{\Omega\,(t-t_f)} - \chi_f \sinh{\Omega (t-t_i)}\right]\ ,
\end{equation}
with $F=0$, where $\chi_i$ is an initial off-equilibrium configuration, and $\chi_f$ is an equilibrium configuration $\chi$ relaxes towards.
The classical contribution to the action is then
\begin{equation}
	S\sub{q}^{cl} = 
	\int_{0}^{t_f} dt\, F(t)\, \chi_{cl}(t)\ ,
\end{equation}
where we take the initial time $t_i=0$.
The quantum contribution is given by 
\begin{equation}\label{eq:quaction}
	S^{qu}\sub{eff} = -i\, \frac{T}{m\,\gamma} \int_{0}^{t_f} \int_{0}^{t_f} dt\,dt'\,F(t) \frac{\delta(t-t')}{\partial_t^2-\Omega^2} F(t')\ ,
\end{equation}
with the associated Green's function
\begin{eqnarray}
	G(t,t') = \frac{2}{\Omega} \frac{e^{\Omega\,(t_i+t_f)}}{e^{2\,\Omega\,t_i}-e^{2\,\Omega\,t_f}} &\times&\left[\,
	\sinh \Omega (t-t_f) \sinh \Omega (t_i-t') \theta(t-t') \right.\nonumber \\
	&+&\left.
	\sinh \Omega (t-t_i) \sinh \Omega (t_f-t') \theta(t'-t)\,
	\right]
\end{eqnarray}
In summary, we arrive at an action for the qubit variables -- hidden in the $F(t)$ -- of the form
\begin{equation}\label{eq:chiaction}
	S\sub{q} = 
	\int_{0}^{t_f} dt\, F(t)\, \chi_{cl}(t)
	-i\, \frac{T}{m\,\gamma} \int_{0}^{t_f} dt\, \int_{0}^{t_f}dt'\, F(t) G(t,t') F(t')\ ,
\end{equation}
describing the evolution of the relevant qubits as the bosonic stochastic variable $\chi$ relaxes -- after a quench described by the boundary conditions $\chi_i$ and $\chi_f$. Note that the second part of~(\ref{eq:chiaction}) is imaginary and this implies that the qubit evolution would be in general non-unitary. This piece involves quartic qubit interactions and would be responsible for scrambling information away as the background evolves stochastically. This is not surprising yet an important observation: we are then able to associate information loss in Hawking radiation to the scheme of coarse-graining over short timescales that results in an effective model of what otherwise is microscopic unitary evolution of information. That is, we see how averaging over chaotic dynamics in Matrix theory is responsible for information loss in the dual low energy M-theory or supergravity. Below, we will see that when this non-unitary piece of the effective dynamics becomes important, we expect the emergence of geometry on the dual M-theory side. Our goal next is to consider scenarios where $\chi$, or $x$ and $\delta x$, are perturbed away from equilibrium, and then we want to track the evolution of the qubits described by $\psi$ and $\delta \psi$.

\subsection{Long timescales}

Consider the qubit couplings given by~(\ref{eq:qubitaction}) where $x$ and $\delta x$ are arranged to start off in an off-equilibrium configuration. Neglecting the back reaction of this perturbation onto the black hole, we can take
\begin{equation}
	\left<X\sub{bh}\right> = \left<x\sub{bh}\right> = \left<\delta X\sub{bh}\right> = 0
\end{equation}
so that we have
\begin{equation}\label{eq:qubitaction2}
	S\sub{ferm} \rightarrow \frac{2\,\sqrt{d}}{(2\,\pi)^{3/2}}\,\frac{R}{\ell\sub{s}^3}\, \left[x\, \overline{\delta\psi}\, \Gamma \delta\psi
	+\delta x\, \overline{\delta\psi}\, \Gamma (\psi-\psi\sub{bh})
	- \delta \psi\, \Gamma (\psi-\psi\sub{bh})\, \overline{\delta x}\right]
\end{equation}
We want to develop the action of the qubits using~(\ref{eq:Seffstoch}), which then gives~(\ref{eq:chiaction}) where $\chi$ represents $x$ or $\delta x$, and $F(t)$ can be read off from~(\ref{eq:qubitaction2}). Before looking at the details, notice that the second term in~(\ref{eq:chiaction}), which is quartic in the qubits, is imaginary and renders the evolution non-unitary. The term is the result of coarse-graining over the stochastic variables $x$ and $\delta x$ and naturally leads to information loss. The scale for this non-unitary piece is
\begin{equation}
	\mbox{non-unitary coupling}\sim \frac{T}{m\,\gamma}\, F^2\,t^2\, G \sim  \frac{T}{m\,\gamma}\left(\frac{R}{\ell\sub{s}^3}\right)^2\, \frac{t}{{\partial_t^2-\Omega^2}}
\end{equation}
given that the propagator $G(t,t')$ scales as $\delta(t-t')/(\partial_t^2-\Omega^2)$ and the fermions are dimensionless. Irrespective of whether $\chi$ represents $x$ or $\delta x$, we have 
\begin{equation}
	\frac{T}{m\,\gamma} \simeq R\ .
\end{equation}
From~(\ref{eq:Vdiagonal}) and~(\ref{eq:Omega}), we have  
\begin{equation}
	\Omega_x \sim V_0 \sim \frac{R}{r\sub{h}^2} = \frac{1}{t\sub{h}}\ ,
\end{equation}
when $\chi$ is identified with $x$;
while from~(\ref{eq:Vdeltax}) and~(\ref{eq:Omega}) we instead have
\begin{equation}
	\Omega_{\delta x} \simeq \frac{g\sub{s}}{\ell\sub{s}} = \frac{1}{t\sub{o}}\ ,
\end{equation} 
when $\chi$ is identified with $\delta x$.
Hence, for $t\lesssim t\sub{o}$, the non-unitary coupling scales as
\begin{equation}
	\mbox{non-unitary coupling}\sim \left(\frac{R}{\ell\sub{s}^2}\right)^3 t^3 \sim \left(\frac{t}{t\sub{o}}\right)^3 = g\sub{eff}(\tau)^2
\end{equation}
whether $\chi$ represents $x$ or $\delta x$.
We then see that this coupling, and hence information loss, sets in for timescales of order $t\sim t\sub{o}$, where the effective dimensionless Yang-Mills coupling becomes order unity and Matrix theory starts describing emergent spacetime geometry in the dual formulation. For shorter timescales, $t\ll t\sub{o}$, the evolution is effectively unitary, given by the first semi-classical term in~(\ref{eq:chiaction}). Note however, that for $t\sub{stoch}< t \ll t\sub{o}$, the dynamics is non-local, given by the Planck scale cluster physics and the light nearest neighbor off-diagonal modes $\delta x$ of the matrices.

\subsection{Short timescales}

Let us first start by writing the full qubit action~(\ref{eq:chiaction}) that follows from using~(\ref{eq:qubitaction2}). 
When $\chi$ is identified with the diagonal coordinate $x$ of~(\ref{eq:matrices}), we have
\begin{equation}\label{eq:effdiagpert}
	S\sub{q} = 
	\int_{0}^{t_f} dt\, f(t)\, x_{cl}(t)
	-i\, \frac{T}{m\,\gamma} \int_{0}^{t_f} dt\, \int_{0}^{t_f}dt'\, f(t) \frac{\delta(t-t')}{\partial_t^2-\Omega^2} f(t')\ ,
\end{equation}
and 
\begin{equation}\label{eq:ft}
	f(t) = -\frac{R}{(2\,\pi)^{3/2}\ell\sub{s}^3}	\sqrt{d} \left(\overline{\delta\psi}\cdot\delta \psi-\overline{\delta\psi}'\cdot\delta\psi'\right)
\end{equation}
obtained from~(\ref{eq:Gamma}) and~(\ref{eq:qubitaction2}), and where $\delta\psi'_\alpha\equiv\delta\psi_{\alpha+8}$ with $\alpha = 1,\ldots,8$. The `dot' represents a sum over 8 qubits, {\em i.e.} $\overline{\delta\psi}\cdot \delta \psi\equiv \sum_\alpha \overline{\delta\psi}_\alpha\delta \psi_\alpha$. As mentioned above, the second non-unitary piece is negligible for $t\ll t\sub{o}$.
Looking at the first term of~(\ref{eq:effdiagpert}), we can see that it provides mass to the $\delta \psi$ and $\delta \psi'$ qubits, and it scales as
\begin{equation}
	x\sub{cl} \frac{R}{\ell\sub{s}^3} t \sim \frac{x\sub{cl}}{\ell\sub{s}} \frac{t}{t\sub{o}}\ .
\end{equation}
For early times where $t<t\sub{o}$, this term is important only if $x\sub{cl}$ is large. This, for example, would be the case if the matrix entry labeled by $x$ would evaporate away, $x\sub{cl} \gtrsim r\sub{h}$. If the initial perturbation for the stochastic variable $x$ is such that $x_i\sim r\sub{h}$, the subsequent stochastic evolution is in a flat potential given the form of~(\ref{eq:potential}). This evolution, described by~(\ref{eq:moment1}) and~(\ref{eq:moment2}) -- or equivalently~(\ref{eq:chicl}), results in $x\sub{cl}(t)$ growing to infinity\footnote{To account for the flat direction in~(\ref{eq:potential}), we can for example take $\Omega_x\rightarrow 0$, which gives from~(\ref{eq:chicl}) $x\sub{cl}(t)\rightarrow x_i+(t/t\sub{o}) (x_f-x_i)$, where $x_i\sim r\sub{h}.$}. We then conclude that the effective qubit dynamics that arises from a perturbation on the diagonal -- that corresponds to $x$ evaporating away -- is described by
\begin{equation}\label{eq:qubits2}
	S^{(I)}\sub{eff} = \int\sub{o}^{t\sub{o}} dt\, f(t) x\sub{cl}(t)\ ,
\end{equation}
with $x\sub{cl} \sim r\sub{h}$ initially and growing larger thereafter. This is the statement that the off-diagonal qubits $\delta \psi$ and $\delta \psi'$ become heavier and heavier and condense as the bit evaporates away.

For the off-diagonal coordinate $\delta x$ in~(\ref{eq:qubitaction2}), the resulting action takes the form
\begin{equation}\label{eq:offdiagqubits}
	S\sub{q} = \int\sub{o}^{t_f} dt\,  (\delta x_{cl}(t) F+\overline{\delta x_{cl}}(t) \overline{F}) - i\,R\, \int\sub{o}^{t_f}dt\int\sub{o}^{t_f} dt'\, F^*(t)\frac{\delta(t-t')}{\partial_{t'}^2-\Omega^2} F(t')\ ,
\end{equation}
where
\begin{equation}
	F(t) = \frac{R}{(2\,\pi)^{3/2}\ell\sub{s}^3} \sqrt{d} \times \left[
	(\psi^+-\psi\sub{bh}^+)\cdot(\delta \psi +i\, \delta\psi')+
	(\psi^--\psi\sub{bh}^-)\cdot(\delta \psi -i\, \delta\psi')\right]
	\label{eq:Ft}
\end{equation}
In arriving at this expression, we have used a complexified version of the action~(\ref{eq:Seffstoch}) where $\chi$ is complex as is $\delta x$ -- since the integrated modes are most naturally represented by complex variables. We also have used the diagonal qubit operators $\psi^\pm$ and $\psi\sub{bh}^\pm$ defined in~(\ref{eq:diagqubits}). 
Once again, as described above, the second non-unitary piece is negligible for $t\ll t\sub{o}$.
The first term in~(\ref{eq:offdiagqubits}) term provides a coupling between qubits $\psi$, $\psi\sub{bh}$, and $\delta\psi$ and it scales as
\begin{equation}
	\delta x\sub{cl} \frac{R}{\ell\sub{s}^3} t \sim \frac{\delta x\sub{cl}}{\ell\sub{s}} \frac{t}{t\sub{o}}\ .
\end{equation}
As the bit $x$ evaporates away, equations~(\ref{eq:moment1}) and~(\ref{eq:moment2}) -- or equivalently~(\ref{eq:chicl}) -- tell us that the initial value of $\delta x\sub{cl}$ decays exponentially to zero on timescale given by $t\sub{o}$, as the mode becomes heavy\footnote{The easiest way to see this is to note that, using~(\ref{eq:moment1}), we have $d(\delta x\sub{cl})/dt = -\Omega_{\delta x} \delta x\sub{cl}$.}. At short times $t\ll t\sub{o}$, we write
\begin{equation}\label{eq:qubits1}
	S^{(II)}\sub{eff} = \frac{1}{2}\int\sub{o}^{t\sub{o}} dt\, (\delta x_{cl}(t)\,F+\overline{\delta x_{cl}}(t)\,\overline{F})\ .
\end{equation}

In summary, the qubit action is given by $S^{(I)}\sub{eff}+S^{(II)}\sub{eff}$, or
\begin{eqnarray}
	S\sub{q} &=& \int\sub{o}^{\tau\sub{o}} d\tau\, \left[\,g\,\xi\sub{cl}(\tau) \left(-\overline{\delta\psi}\cdot\delta \psi+\overline{\delta\psi}'\cdot\delta\psi'\right) \right. \nonumber \\
	&+& \left. g\, \delta \xi_{cl}(\tau)\,\left(
	(\psi^+-\psi\sub{bh}^+)\cdot(\delta \psi +i\, \delta\psi')-
	(\delta \psi -i\, \delta\psi')\cdot(\psi^--\psi\sub{bh}^-)\right) \right. \nonumber \\
	&+& \left. g\,\overline{\delta \xi_{cl}}(\tau)\,\left(
	(\overline{\delta \psi} - i\, \overline{\delta\psi'})\cdot(\psi^--\psi\sub{bh}^-)-
	(\psi^+-\psi\sub{bh}^+)\cdot(\overline{\delta \psi} + i\, \overline{\delta\psi'})\right)\ ,
	\right]\label{eq:qubitevolutionaction}
\end{eqnarray}
where we have switch from time $t$, $x$, and $\delta x$ to scaled variables $\tau$, $\xi$, and $\delta \xi$ (see equation~(\ref{eq:rescaled})), and the effective coupling $g$ is defined as
\begin{equation}
	g = \frac{\ell\sub{s} g\sub{s}^{-2/3}R}{(2\,\pi)^{3/2}\ell\sub{s}^3}\sqrt{d} = \frac{(g\sub{YM}^2)^{1/3}\sqrt{d}}{(2\,\pi)^{3/2}}
\end{equation}
which has units of length such that $g\,\tau$ is the effective dimensionless coupling.  
In total, the system describes $8\times 4$ qubits: $8\times 2$ off-diagonals ones denoted by $\delta\psi_\alpha$ and $\delta\psi'_\alpha$, and $8\times 2$ on the diagonal denoted by $\psi_\alpha$ and ${\psi\sub{bh}}_\alpha$. The stochastic relaxation from a quench is given by the classical profiles $\xi\sub{cl}(\tau)=x\sub{cl}(\tau)/\ell\sub{s}$ and $\delta \xi\sub{cl}(\tau)=\delta x\sub{cl}(\tau)/\ell\sub{s}$ that follow from~(\ref{eq:chicl}).

We now elaborate on the implications of the qubit evolution action~(\ref{eq:qubitevolutionaction}), restricting our attention to early times $t\sub{stoch}<t\lesssim t\sub{o}$ -- before the onset of dissipation and emergence of geometry. For the remaining discussion, we will use the coherent state representation of the qubits, which we first briefly review. For a qubit with states $\left|0\right>$ and $\left|1\right>$, a representation over a coherent state $\left|\eta\right>$ looks like~\cite{Grosche:1998yu}
\begin{equation}
	\left<\eta|0\right> = 1\ \ \ ,\ \ \ 
	\left<\eta|1\right> = \overline{\eta}\ ,
\end{equation}
where $\eta$ is a Grassmanian.
A general state $\left|\Phi\right>$ is then a function over the Grassmanians
$\left<\eta|\Phi\right> \equiv \Phi(\overline{\eta})$. A Bell state
\begin{equation}
	\left|B\right> = \frac{1}{\sqrt{2}} \left(\left|0\right>\left|1\right>-\left|1\right>\left|0\right>\right)
\end{equation}
is then represented as
\begin{equation}
	\left<\eta_1\eta_2|B\right> = \frac{1}{\sqrt{2}} \left(\overline{\eta}_2-\overline{\eta}_1\right)\ .
\end{equation} 
The expectation value of an operator gets a form of a function over Grassmanians
\begin{equation}
	\mathcal{O}(\overline{\eta},\eta') \equiv \left<\eta | \mathcal{O}| \eta'\right> = \sum_{m,n=0,1} \overline{\eta}^m \left<m|\mathcal{O}|n\right> {\eta'}^{n}\ .
\end{equation}
The path integral measure is such that
\begin{equation}\label{eq:application}
	\left<\eta|\mathcal{O}|\Phi\right> = \int d\overline{\eta}'d\eta' e^{-\overline{\eta}'\eta'} \mathcal{O}(\overline{\eta},\eta') \Phi(\overline{\eta}')\ .
\end{equation}
For a Hamiltonian of qubits referenced by the operators $\psi^\pm$, we would write $\psi^+\rightarrow \overline{\eta}$ and $\psi^-\rightarrow \eta$. For a simple bilinear and time-dependent structure with sources, we have
\begin{equation}\label{eq:Hform}
	H = A(t)\, \overline{\eta}\eta - \overline{J}(t) \eta - \overline{\eta} J(t)\ .
\end{equation}
The unitary evolution operator as a function over Grassmanians takes the form
\begin{equation}
	U({\overline{\eta}}'', \eta'; t'',t) = \int_{\eta(t')=\eta'}^{\overline{\eta}(t'')=\overline{\eta}''} \mathcal{D}\overline{\eta}(t)\mathcal{D}\eta(t)
	\exp\left[
	\overline{\eta}''\eta(t'') +i\,\int_{t'}^{t''} dt\,\left(i\,\overline{\eta}(t) \dot{\eta}(t)-H(\overline{\eta}(t),\eta(t),t)\right)
	\right]\ ,
\end{equation}
which, for a Hamiltonian of the form~(\ref{eq:Hform}), then leads to 
\begin{eqnarray}
	U({\overline{\eta}}'', \eta'; t'',t) &=& 
	\exp\left[
	{\overline{\eta}}''e^{-i \int_{t'}^{t''}A(t) dt} \eta' \right. \nonumber \\
	&+&\left.i\, {\overline{\eta}}'' \int_{t'}^{t''} dt\, J(t) e^{-i \int_{t'}^{t''} ds\,\theta(s-t) A(s)}
	+i\, \int_{t'}^{t''} dt\, \overline{J}(t)e^{-i\int_{t'}^{t''}ds\,\theta(t-s)A(s)}\eta' \right. \nonumber \\
	&-&\left. \int_{t'}^{t''} dt \int_{t'}^{t''} ds\, \overline{J}(t)D(t,s)J(s)
	\right]\label{eq:master}
\end{eqnarray}
where the propagator is given by 
\begin{equation}
	D(t,s) \equiv \theta(t-s) e^{i \int_{s}^t A(t') dt'}\ .
\end{equation}
We can then use this approach to write the unitary evolution operator for the qubits given by~(\ref{eq:qubitevolutionaction}). The Grassmanian variables will be labeled as $\delta\psi$, $\delta\psi'$, $\psi^-$, $\psi\sub{bh}^-$, and their complex conjugates  -- in correspondence with the respective operators. We then seek the evolution operator written as 
\begin{equation}\label{eq:U}
	U(\psi^+(\tau), \psi^-(0), \psi\sub{bh}^+(\tau), \psi\sub{bh}^-(0), \overline{\delta\psi}(\tau),  {\delta\psi}(0), \overline{\delta\psi'}(\tau),  {\delta\psi'}(0) ;\tau, 0)	
\end{equation} 
that acts on the qubit wavefunction $\Phi(\psi^+(0), \psi\sub{bh}^+(0), \overline{\delta\psi}(0),\overline{\delta\psi'}(0))$. We have the evolution of a $8\times 4$ qubit system, half on the matrix diagonal and the other half off-diagonal; all $32$ qubit are part of a cluster. The time evolution is obviously sensitive to the details of the quench, given by $x\sub{cl}(t)$ and $\delta x\sub{cl}(t)$. The initial wavefunction $\Phi$ is another input to the problem. Cluster formation dynamics might naturally involve the delicate physics of D0 bound state formation -- akin to Cooper pair formation in superconductivity. The dynamics of the marginal bound states in Matrix theory is a complicated strong coupling problem that remains an open issue, and we will not be able to tackle the full problem here. Instead, given the spirit of an effective approximate scaling analysis, we will next engage in a speculative analysis that is inspired by a recent toy model of black hole qubit evaporation due to Osuga and Page~\cite{Osuga:2016htn}. We will argue that the Matrix theory qubit evolution operator has the hallmarks of the toy model presented in~\cite{Osuga:2016htn}, under a series of assumptions. 

In~\cite{Osuga:2016htn}, a toy model was proposed whereby the black hole Hilbert space is augmented to a tensor product that involves the black hole qubit sector and two other sectors, one for in-falling and another for outgoing radiation modes just inside and just outside the event horizon. Each black hole qubit is paired with two qubits that are in the singlet Bell state. The latter is proposed to represent the vacuum for the radiation pair of modes that assures smooth spacetime near the horizon. As a black hole qubit evaporates away, \cite{Osuga:2016htn} proposes a unitary evolution operator that essentially exchanges the black hole qubit with a qubit of outgoing radiation, leaving the black hole sector qubit entangled in a Bell state with the qubit of incoming radiation. The result of this is that one qubit of information leaves the black hole (into the outgoing radiation sector), and a vacuum Bell state of two qubits (black hole and incoming radiation sectors) is left behind that is now to be interpreted as part of a bit of new empty spacetime created just outside a black hole as the latter shrinks in size. The key assumptions in this model are: interactions in the black hole
qubit sector are non-local at the Planck scale, and a Bell vacuum state for black hole and incoming radiation qubits is tantamount to shrinking the black hole or equivalently expanding the vacuum space outside of it. The motivation for this toy example is to present a proof of concept model of black hole evaporation consistent with black hole complementarity.

In our setup, we have an explicit quantum theory of gravity that dictates the qubit evolution operator. The partons of the matrix black holes are clusters of diagonal and off-diagonal matrix qubits, about $8\times (d-1)^2$ qubits in $d$ space dimensions. For $d=3$, that's $32$ qubits. We propose that each cluster of qubits, a $32$-qubit system, carries $8$ qubits worth of information only -- corresponding to the $256$ supergravity states that can encode information; the remaining $24$ qubits are scaffolding that are in a highly entangled Bell-like vacuum state that is the result of cluster dynamics. These represent the halo at around the event horizon. Naturally, the information is on the diagonal qubits, say in $\psi\sub{bh}$ in the specific setup we have been considering. That means that $\delta\psi$, $\delta \psi'$, and $\psi$ start off in a maximally entangled vacuum Bell state of $24$ qubits representing radiation or `membrane goo' near the horizon. We then propose that the unitary evolution operator from~(\ref{eq:U}) and~(\ref{eq:qubitevolutionaction}) -- given a perturbation of the stochastic variables $x\sub{cl}(t)$ and $\delta x\sub{cl}(t)$ that describes the evaporation of the $x$ matrix entry -- results in having the qubit of information $\psi\sub{bh}$ transfered to $\psi$ which exits the Matrix black hole. The end result leaves behind a vacuum Bell state of qubits for $\delta\psi$, $\delta \psi'$, and $\psi\sub{bh}$ that is to be interpreted as the production of a bit of new spacetime outside the black hole. As a result, the matrix black hole shrinks in size from $N$ to $N-2$. Looking at the form of~(\ref{eq:qubitevolutionaction}), we see a structure that has the right general form to potentially generate such an evolution of qubits. The analogue of the exchange operator from~\cite{Osuga:2016htn} in our language takes the form $\exp\left[i\,\alpha\,t \,(\psi^+-\psi\sub{bh}^+)(\psi^--\psi\sub{bh}^-)\right]$. Our effective Hamiltonian involves in addition the mediation of the light $\delta\psi$ modes in combinations of the form $\sim(\psi^+-\psi\sub{bh}^+)\delta \psi^-$ and its complex conjugate.

Bell states with $24$ qubits are very difficult to study and even determine in their own right. Added to this complication is the fact that~(\ref{eq:qubitevolutionaction}) is in general non-local due to the light off-diagonal modes. As a result, it is a very challenging task to determine the evolution of the qubits using the action~(\ref{eq:qubitevolutionaction}). To see this, note that the non-local couplings in~(\ref{eq:master}) have scale given by
\begin{equation}
	\int_0^{\tau_0} d\tau' g\, \chi\sub{cl}(\tau) \simeq (\chi_i+\chi_f)\times g\, \tau_0 \sim (\chi_i+\chi_f)
\end{equation}
where we used~(\ref{eq:chicl}). For $\chi\rightarrow \xi\sub{cl} \gg 1$, given that $r\sub{bh}\gg \ell\sub{s}$.  For $\chi\rightarrow \delta\xi\sub{cl} \sim 1$, given that the cluster length scale is $\ell\sub{s}$. In any scenario, the relevant dynamics is highly non-local. Noting some of the general similarities between the model of~\cite{Osuga:2016htn} and ours, we leave the analysis of the significantly more complex dynamics of our system for future work.

\section{Discussion and Outlook}\label{sec:conclusion}
\label{sub:conclusion}

The analysis in this work is a first attempt to develop a quantum gravity-centric, bottom up  picture of black hole event horizon physics. The results can be summarized through two main conclusions:

\begin{enumerate}
	\item We have determined that near horizon dynamics is non-local in space {\em and} time at the Planck scale. The thermal degrees of freedom of the black hole are `cells' of around $d$ particles, for a black hole in $d$ space dimensions; each cell spans a size of order the Planck scale. One can think of each cell carrying bits of information, encoded in the polarization states of the fermionic variables of Matrix theory -- or equivalently the polarization states of the supergravity multiplet on the dual side. The dynamics of black hole degrees of freedom is non-local and chaotic for short Planckian timescales, in a regime where the Yang-Mills theory is hovering just below strong coupling. At longer timescales and larger distances, the dynamics is effectively local both in time and space, while being strongly coupled. This is when and where an effective geometrical picture is possible.
	\item When describing evaporation, one is dealing with a chaotic system near the would-be event horizon with a characteristic timescale given by the Planck scale. To describe the evaporation via a top down approach, {\em i.e.} via Hawking's approach, one needs to average chaotic dynamics over super-Planckian timescales. Where a spacetime description is valid, one is necessarily left with a non-unitary effective picture for the evaporation arising from coarse graining over Planckian chaotic motion. The suggestion is that the resolution of the black hole information loss paradox {\em cannot} lie in any framework that relies on a well-defined smooth spacetime geometry at the event horizon. This is a plausibility argument: We demonstrated that, through a rather simple stochastic model with a single input scale, one can understand how Hawking evoporation is inherently non-unitary -- naturally due to stochastic, chaotic UV physics. This simplest of settings necessitates however the breakdown of smooth geometry at the horizon. This obervation, together with other independent evidence towards a breakdown of geometry at the horizon, constitute strong evidence that one most likely needs to look for resolutions of the information paradox in models involving a new perspective on near horizon geometry. The geometrical description of black hole evaporation is inherently non-unitary as it arises from averaging over Planckian timescales that characterize the chaotic physics of the underlying degrees of freedom.
\end{enumerate}

A couple of footnotes are in order. First, we identify emergent geometry at the benchmark of strong effective Yang-Mills coupling $g\sub{eff}(\tau)^2$, as opposed to strong effective {\em 't Hooft} coupling $g\sub{eff}(\tau)^2 N$, which is the natural coupling for large $N$. The subtlety here is that the coupling that governs the microscopic event horizon dynamics is one that arises from the interaction of `order one' matrix entries on the diagonal. At most groups of order $d^2$ particles participate in the dynamics, hence the relevant effective coupling is not the $N$ dependent 't Hooft coupling. In describing the gravitational interaction of the {\em whole} black hole with entropy $S\sim N$, the relevant effective coupling is indeed the 't Hooft coupling; but microscopic event horizon dynamics does not involve the participation of all $N$ degrees of freedom.

The second footnote has to do with implicit connections to the issue of black hole complementarity~\cite{Susskind:1993if}-\cite{Susskind:2012rm} . In modeling the mean field potential for the degrees of freedom of the Matrix black hole, we note that there was no need to introduce a separate Planck scale near the horizon: the entire potential can be modeled using a 
single scale, the radius of the event horizon\footnote{Our model builds from the outset on the premise that Hawking evaporation is a single scale phenomenon, at least to leading order. This does not allow capturing new UV physics through this model that might still exist  and correct Hawking evaporation. Yet, the point is that such additional scales are not needed to understand why Hawking evporation is inherently non-unitary.}. This is not surprising since we were modeling the physics in a manner to match against expectations on the dual supergravity side. We also noted that the qubit action we arrived at has some of the features of the qubit evolution toy model proposed in the work of Osuga and Page~\cite{Osuga:2016htn}. The latter consisted of a proof-of-concept system that circumvents the need of a firewall by positing non-local interactions at the horizon and an exchange mechanism of qubits within a direct product of three Hilbert spaces. All these ingredients of this toy model emerge naturally from our Matrix theory discussion. However, our action is more complicated than the one in~\cite{Osuga:2016htn}, and we leave a detailed analysis of the dynamics for future work. Nevertheless, these similarities between the two systems, ours and that of~\cite{Osuga:2016htn}, might be hints that a firewall is {\em not} needed at the event horizon after all, and black hole complementarity prevails. This is consistent with~\cite{Giddings:2011ks,Giddings:2012bm,Almheiri:2012rt} given the non-local nature of the interactions near the event horizon in Matrix theory -- at the level of 
D0 brane clusters. There is however a significant conceptual challenge to this argument. Black hole complementarity is a statement about the perspective of an in-falling observer. This means that one needs to understand how a change of perspective between the observer at infinity and the one in-falling past the horizon is realized in the language of Matrix theory. Presumably, this involves a Matrix transformation in $U(N)$ since one expects that local spacetime coordinate invariance is embedded in the gauge group of the theory. This in turn requires a more precise map between emergent geometry and metric, and matrix degrees of freedom. Without this critical missing ingredient, we cannot conclusively understand how the firewall paradox is addressed by our effective model.

Related to this last point, we also note that our treatment explicitly chooses a frame for describing the black hole, presumably corresponding to the perspective of an outside observer. This creates a clear separation between the roles of diagonal and off-diagonal matrix entries. The residual gauge freedom is the group of permuting diagonal entries, a subgroup of $U(N)$. The more interesting transformations would mix diagonal and off-diagonal entries, and we believe these correspond in part to switching the perspective of the observer. Very little is known or understood about this part of the Matrix-supergravity duality, and it seems a full treatment of the quantum black hole would necessitate progress in this direction.

This work is a step towards unravelling the microscopic details of black hole horizon physics within a theory of quantum gravity that is fully embedded in string/M-theory. The effective model approach opens up new directions for a range of possible investigations and extensions that can only add to our understanding of black holes and quantum gravity. We hope to report on some of these in future works.

\newpage
\section{Acknowledgments}

This work was supported by NSF grant number PHY-0968726.

\providecommand{\href}[2]{#2}\begingroup\raggedright\endgroup

%

\end{document}